\documentclass[%
reprint,
superscriptaddress,
amsmath,amssymb,
aps,
]{revtex4-2}

\usepackage{graphicx}
\usepackage{dcolumn}
\usepackage{bm}
\usepackage{braket}
\usepackage{comment}
\usepackage{color}
\usepackage{verbatim}
\usepackage{dsfont}
\usepackage{placeins}
\usepackage{soul}
\usepackage[rgb]{xcolor} 
\usepackage{import}
\usepackage{algpseudocode}

\newcommand{\Sij}[3]{ \left \langle #1 \right | #2 \left | #3 \right \rangle}

\newcommand{\tcr}{\textcolor{red}}

\newcommand{\mean}[1]{\ensuremath{\left\langle #1 \right\rangle}}

\renewcommand{\st}[1]{~}
\renewcommand{\tcr}[1]{~}

\begin{document}

\preprint{APS/123-QED}

\title{Clifford Circuit Initialisation for Variational Quantum Algorithms}

\author{M. H. Cheng}
\email{mhc2617@ic.ac.uk}
\affiliation{QOLS, Blackett Laboratory, Imperial College London, London, SW7 2AZ, UK\\
}
\affiliation{
	Fraunhofer Institute for Industrial Mathematics, Fraunhofer-Platz 1, 67663 Kaiserslautern, Germany\\
}

\author{K. E. Khosla}
\email{k.khosla@imperial.ac.uk \\ These two authors contributed equally to this work}
\affiliation{QOLS, Blackett Laboratory, Imperial College London, London, SW7 2AZ, UK\\
}

\author{C. N. Self}
\affiliation{
	Quantinuum, Partnership House, Carlisle Place, London, SW1P 1BX, United Kingdom\\
}
\affiliation{QOLS, Blackett Laboratory, Imperial College London, London, SW7 2AZ, UK\\
}

\author{M. Lin}
\affiliation{QOLS, Blackett Laboratory, Imperial College London, London, SW7 2AZ, UK\\
}

\author{B. X. Li}
\affiliation{QOLS, Blackett Laboratory, Imperial College London, London, SW7 2AZ, UK\\
}

\author{A. C. Medina}%
\affiliation{
	Fraunhofer Institute for Industrial Mathematics, Fraunhofer-Platz 1, 67663 Kaiserslautern, Germany\\
}%

\author{M. S. Kim}
\affiliation{QOLS, Blackett Laboratory, Imperial College London, London, SW7 2AZ, UK\\
}




\date{\today}

\begin{abstract}
	We present an initialisation method for variational quantum algorithms applicable to intermediate scale quantum computers. The method uses simulated annealing of the efficiently simulable Clifford parameter points as a pre-optimisation to find a low energy initial condition. We numerically demonstrate the effectiveness of the technique, and how it depends on Hamiltonian structure, number of qubits and circuit depth. While a range of different problems are considered, we note that the method is particularly useful for quantum chemistry problems. This presented method could help achieve a quantum advantage in noisy or fault-tolerant intermediate scale devices, even though we prove in general that the method is not arbitrarily scalable. 
\end{abstract}

\maketitle

\section{Introduction}
Intermediate Scale quantum devices are rapidly improving and are potential candidates for demonstrating quantum advantage. While algorithms such as the phase estimation~\cite{phase_est}, factorisation~\cite{Shor}, or Deutsch-Jozsa~\cite{Deutsch} are too resource intensive to demonstrate quantum advantage on current devices, intermediate scale devices have inspired an important class of hybrid classical-quantum algorithms based on classical optimisation – Variational Quantum Algorithms (VQA)\cite{VQE_1, VQE_2, VQE_3}. Device noise is a significant barrier to achieving supremacy with VQAs on noisy Intermediate Scale Quantum (noisy ISQ) devices~\cite{Noise_limitation}; however, due to their ability to efficiently estimate classically intractable cost functions, VQAs remain a possible candidate for demonstrating quantum advantage on small, or intermediate scale error corrected devices. This class of algorithms divides an eigenvalue problem into a classical optimisation task, using a quantum computer to estimate the cost function. The calculation of the cost function requires only a relatively short quantum circuit, replacing the requirement for deep circuits and ancillary qubits with a run-time overhead. The Variational Quantum Eigensolver (VQE), is a specific variational algorithm that aims to minimize the expectation value of a given Hamiltonian. The VQE is sufficiently flexible to solve many optimization problems, including MAXCUT~\cite{QAOA,amaro_filtering_2022}, portfolio optimisation~\cite{Profolio_Opt}, financial transaction settlement~\cite{financial_settlement}, and finding the ground state of solid-state Hamiltonians~\cite{solid-state} or interacting fermions~\cite{SU(N)_fermions}.

The VQE has had some initial success in demonstrating simulations of quantum chemistry energy levels within the chemical accuracy~\cite{Hartree-Fock}, exotic phases of matter such as time crystals~\cite{Time_Crystal}, and high energy physics~\cite{high_energy}. However, scalability remains a concern as the numbers of qubits and parameters increase~\cite{Scalability}, notwithstanding the issue with noise~\cite{Noise_limitation}. The main obstacles to scalability are the onset of barren plateaus~\cite{barren_plateau, entanglement_barren} and the trade-off between expressibility and trainability~\cite{Holmes_2022} of a parameterized quantum circuit.

To tackle these problems, we propose a classical initialization strategy for the VQE: the \textit{Clifford Circuit Pre-Optimisation}, which classically searches for a good initial condition before running the VQE on device. Clifford circuits are quantum operations that map Pauli strings to Pauli strings, and when applied to an initial stabilizer state (i.e. an eigenstate of a Pauli string), the resulting measurement outcomes are known to be efficiently simulable~\cite{Aaronson_2004}. This pre-optimisation can help avoid barren plateaus and remove the influence of device noise during the classical optimisation. Given an ansatz circuit consisting of parameterized $R_z$, $R_x$, and $R_y$ rotations as well as CNOTs, there exists a finite set of parameters (Clifford points) such that the final quantum state is a stabilizer state and Pauli measurement outcomes are classically simulable. Since stabilizer states are uniformly distributed in the Hilbert space, one can envisage a search algorithm over these discrete points to seek a good initialization. In this work, we consider classical annealing with threshold resetting as the search method, however, in writing the current manuscript, we became aware of a complimentary work using Bayesian optimisation with encouraging results~\cite{CAFQA}. 

A good initialization needs not just to minimize the cost function; ideally we would like an initialization with both low cost, and high gradient. This gradient can be efficiently evaluated using the parameter-shift rule~\cite{parametershift}, which itself also only requires Clifford circuits for a suitable choice of parameters. 

In this paper, we initially study the advantages of the Clifford pre-optimization by considering the initial condition with low cost before running the continuous parameter variational optimisation. We numerically show how Clifford pre-optimisation works with different Hamiltonian models, and that starting at a low-cost initial condition drastically reduces the gradient descent times. In section~\ref{section: basics} we overview the Clifford pre-optimisation method, comparing and contrasting it to standard continuous parameter VQE. We then present numerical results in section~\ref{section: results} for quantum chemistry, Transverse Field Ising Model (TFIM), and Binary optimisation problems using the Quantum Approximate Optimisation Algorithm (QAOA). We also investigate the relationship between ansatz types, circuit depth and numerical scalability. Finally, in section~\ref{section:scaling}, we use a counting argument to show the method is not arbitrarily scalable, but is still relevant for pushing true quantum advantage for ISQ circuit sizes of several tens of qubits.

\section{Clifford pre-optimisation}
\label{section: basics}

\begin{figure}
	\centering
	\includegraphics[width = 0.95\columnwidth]{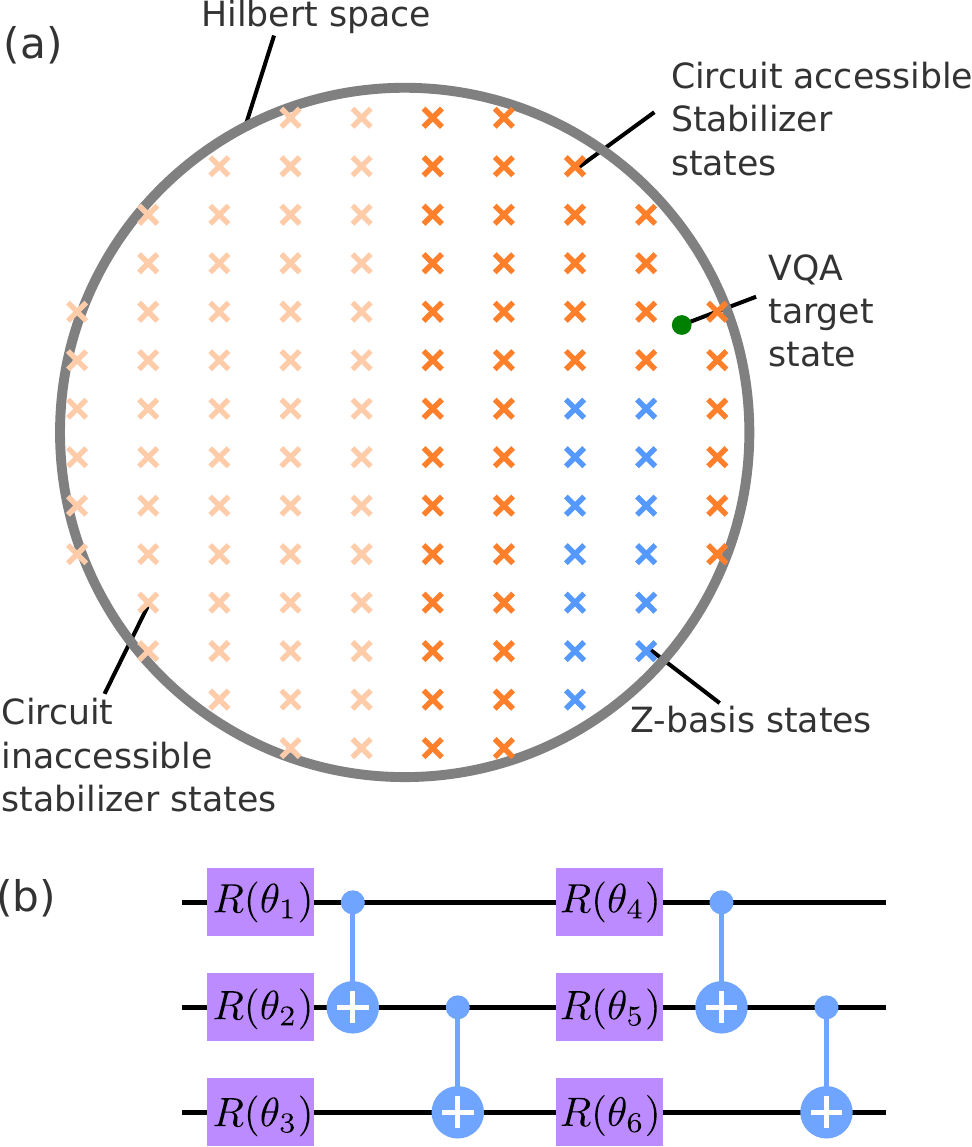}
	\caption{(a) Schematic representation of Clifford searching for the VQE, with crosses depicting the stabilizer states in the Hilbert space. A circuit typically parameterizes a subset (darker crosses) of all stabilizer states  --- generally including the logical Z-basis states (blue crosses). A cost function can be efficiently classically simulated at these points, and a pre-optimisation method can be used to find a state close to the VQE target state (green dot). (b) A hardware efficient ansatz can become a Clifford circuit by restricting ($\theta_i$)-parameterised rotations to a finite set, shown here as single direction Pauli rotations with angles being multiples of $\pi/2$.}
	\label{fig:schematic} 
\end{figure}
The VQE algorithm requires a parameterised ansatz circuit $U(\theta)$ to construct the trial wave function $\ket{\psi(\theta)}$, combined with a classical optimisation that minimizes the expectation value $\Sij{\psi(\theta)}{H}{\psi(\theta)}$ for some operator $H$ (which is decomposed into a weighted sum of Pauli strings). Selecting the structure and initial parameters for this ansatz circuit is an active area of VQE research~\cite{Initialisation_1, Initialisation_2, Initialisation_3, entanglement_barren}. For typical ansatz circuits, there is a finite set of parameter points that implement a Clifford circuit, with the resulting states being the so-called stabilizer states, Fig.~\ref{fig:schematic} (a). In the following we make use of the Gottesman-Knill theorem~\cite{Gottesman} to efficiently classically sample the cost function (over these Clifford circuit-parameterized stabilizer states) before any quantum hardware is needed. A hardware-efficient Clifford circuit can be constructed by restricting the parameter angles in single-direction rotation gates, Fig.~\ref{fig:schematic} (b). Alternatively, one could consider a regular circuit of random single qubit Clifford gates between entangling C-NOT (or other two-qubit Clifford) layers, and later decide which possible angles to relax for a continuous variable circuit.

For some problems the Clifford points of the relevant ansatz may not be the physically interesting part of the variational parameter space, e.g. small perturbations using the Universal Coupled Cluster ansatz \cite{UCCSD}. For the present work, we restrict our analysis to the problem agnostic ans\"{a}tze, where the physically relevant part of the variational parameter space is unknown. Before presenting numerical results, we first consider in which case this Clifford pre-optimisation search may be useful and identify potential limitations. 

Implementations of continuous parameter VQE are known to have a trade-off between expressibility and trainability~\cite{Holmes_2022}. The same problem is expected in the stabilizer pre-optimisation: (i) the ansatz may not be sufficiently expressive to include the best stabilizer state (i.e. the stabilizer state that minimises the energy), or (ii) the circuit can express too many stabilizer states and the pre-optimisation itself becomes difficult. If there are too many parameters, the finite-size ans\"{a}tze begin to approximate a 2-design~\cite{2-design}, where barren plateaus and uniform cost function conditions apply~\cite{barren_plateau}. Thus, the right number of parameters and circuit design will be the key to a successful Clifford pre-optimisation.

For quantum chemistry tasks, a Fermion-to-qubit transform, e.g. Jordan-Wigner~\cite{Jordan-Wigner}, Parity~\cite{Parity}, or Bravyi-Kitaev~\cite{Bravyi}, is required to generate a suitable qubit Hamiltonian. This transform has a direct impact on ``where'' (in Hilbert space) the ground state is located, in contrast to parametrically accessible stabilizer states which are fixed in the Hilbert space. Hence different Fermion-to-qubit transforms may be more or less suitable for pre-optimisation given a particular hardware-efficient ansatz. Unfortunately, it is not possible to \emph{a priori} know the best transform, but running the stabilizer search over each transform is a linear overhead and the best transform for the quantum device can therefore be selected in a scalable way.

\begin{figure*}
	\centering
	\includegraphics[width=.9\textwidth]{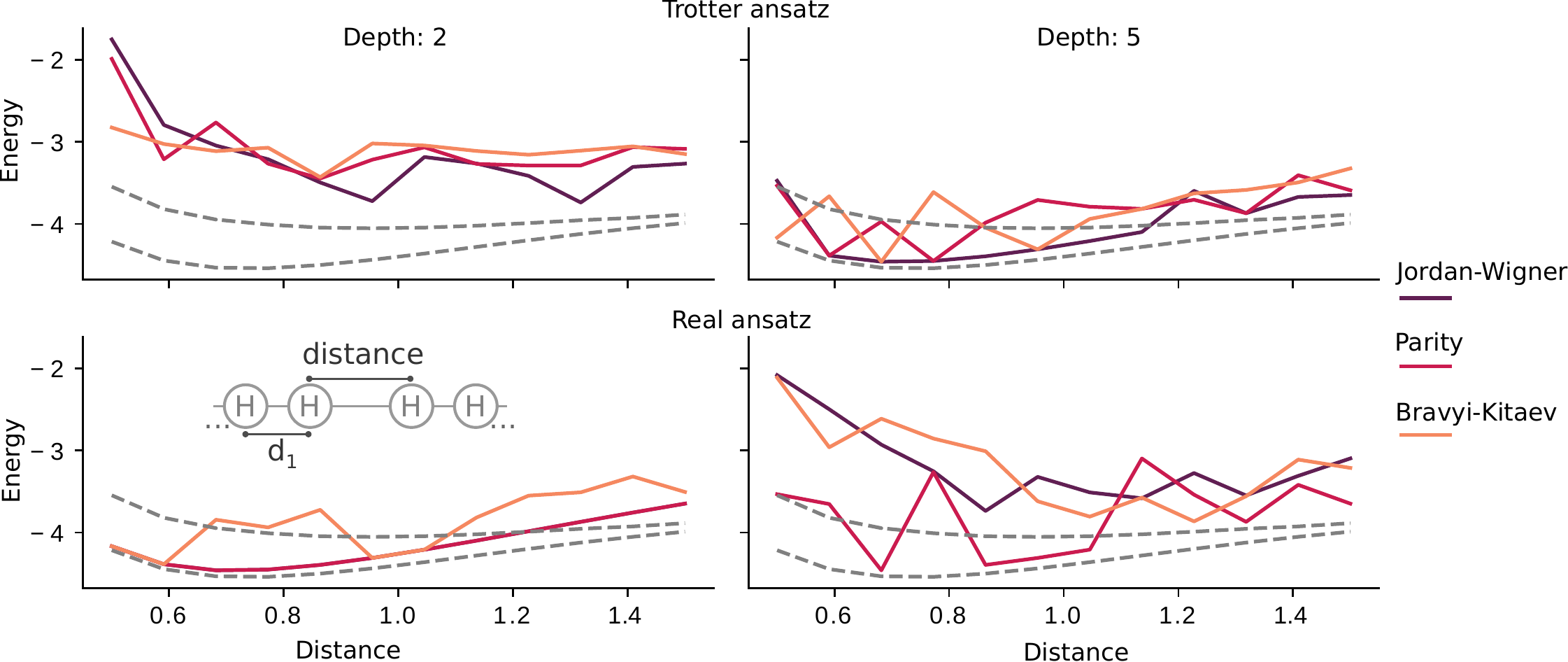}%
	\caption{Comparison of different ansatz types (rows), ansatz depths (columns), and Fermion transforms (line colours) for a linear chain of eight hydrogen atoms (see inset). The energy curve is obtained from ten thousand iterations of Clifford pre-optimisation, with no quantum VQE. Grey dashed lines show ground and first excited states obtained via exact diagonalization. The real-valued ansatz of depth two, and Trotter ansatz of depth five perform equally well at approximating the ground state energy for low bond lengths. The Trotter ansatz at depth two is insufficiently expressive, while the real-valued ansatz at depth five is over parameterized. The plots highlight how the pre-optimisation is not a panacea to all problems, and the ansatz structure, depth and Fermion-to-qubit transform each play an interesting role. Further on-device VQE is necessary to find better energy solutions (e.g. up to chemical accuracy). We note that these curves are found classically and can therefore be used to select the best ansatz for on-device optimisation.}
	\label{fig:H8}
\end{figure*}

Finally, we note that a given Pauli string has far more $0$ expectation value stabilizer states than $\pm 1$ eigenvalue stabilizers, and therefore the expectation value of a Pauli string with a random stabilizer state is likely zero. These zero expectation value states give no information to the optimiser resulting in a barren plateau-like flattening of the optimisation landscape. Higher numbers of non-commuting Pauli strings will help combat this, hence the more Pauli strings, the more likely a given stabilizer state will give you useful information. These arguments are expanded in detail in section~\ref{section:scaling}, but for now we simply postulate that Hamiltonians with more (non-Commuting) Pauli strings may be better suited to Clifford pre-optimisation. Hence the Clifford pre-optimisation cannot be expected to be useful for all problems at every scale. Since quantum chemistry problems typically require many non-commuting Pauli strings, we might expect that the stabilizer search is more useful for chemistry problems compared to binary optimisation (e.g. QAOA), or quadratic condensed matter interactions.

\section{Numerical results}
\label{section: results}
Here we consider several example problems to test the pre-optimisation method. Different Hamiltonians are chosen to span local (MAXCUT and Ising model) and global (quantum chemistry examples) cost functions. Here the distinction between local and global loosely refers to how many non-identity Pauli operators are in each of the Pauli strings in the decomposition of the Hamiltonian~\cite{k-local}. Furthermore, we investigate the effect of ansatz circuit type, circuit depth, and Fermion-to-qubit transformations (for quantum chemistry problems) on the pre-optimisation. In each case, simulated annealing with a threshold resetting method is used to optimize each Clifford circuit. Our annealing method begins with random Clifford parameters and changes two Clifford gates each iteration, keeping the new gates if a lower energy state is found, and probabilistically keeping the new gate if a higher energy state is found (see appendix~\ref{section: numericals} for details). 

First, as a typical benchmark, we consider ground state problems in quantum chemistry. We use the \textit{qiskit\_nature} library to compute the molecular orbitals of a second quantized Fermionic Hamiltonian with frozen core electrons (if applicable). The same library is used to generate a qubit Hamiltonian via the parity~\cite{Parity}, Jordan-Wigner~\cite{Jordan-Wigner}, or Bravyi-Kitaev~\cite{Bravyi} transforms. These quantum chemistry Hamiltonians are expressed as a weighted sum of Pauli strings, with many groups of non-commuting terms.

Following on from quantum chemistry problems, we then analyze Clifford pre-optimization as it applies to the MAXCUT and Transverse Field Ising Model (TFIM) Hamiltonians. Both of these Hamiltonians can be expressed as, 
\begin{eqnarray}
	H = \sum_{ij}w_{ij}Z_iZ_j + \sum_i (g_x X_i + g_z Z_i).
	\label{eq:H}
\end{eqnarray}
For MAXCUT, $w_{ij}$ is the graph adjacency matrix and $g_x=g_z = 0$, while for the TFIM $w_{ij} = J \delta_{i,i+1}$ with coupling constant $J$ (and Kronecker-delta $\delta_{ij}$), and $g_x$ ($g_z$) the transverse (longitudinal) field. Other constrained and unconstrained quadratic binary optimisation problems can be written in this Ising-like form~\cite{QUBO_Ising, financial_settlement}. While the TFIM Hamiltonian is a local cost function Hamiltonian, it is known to have a phase transition where the ground state exhibits volume law entanglement~\cite{Volume_law}, allowing the potential for a non-trivial stabilizer state to be found via the Clifford search.

As well as different types of Hamiltonians and transforms, we compare the performance of different types of ansatz: (i) an ansatz that parameterizes only real-valued stabilizer states, consisting of only controlled $Y$ rotations and CNOT gates, (ii) an ansatz inspired by the Trotterized unitary for the TFIM Hamiltonian that allows complex-valued stabilizer states, and (iii) for the MAXCUT task we consider the standard QAOA ansatz. The real-valued and QAOA ans\"{a}tze are more constrained than the trotterized ansatz and parameterise fewer stabilizer states for a given depth, but may be more suited to problems with known real-valued ground state eigenvectors, such as some quantum chemistry problems and binary optimisation. It turns out however, that this is not necessarily the case, as we shall later show.

\begin{figure*}
	\centering
	\includegraphics[width=0.8\textwidth]{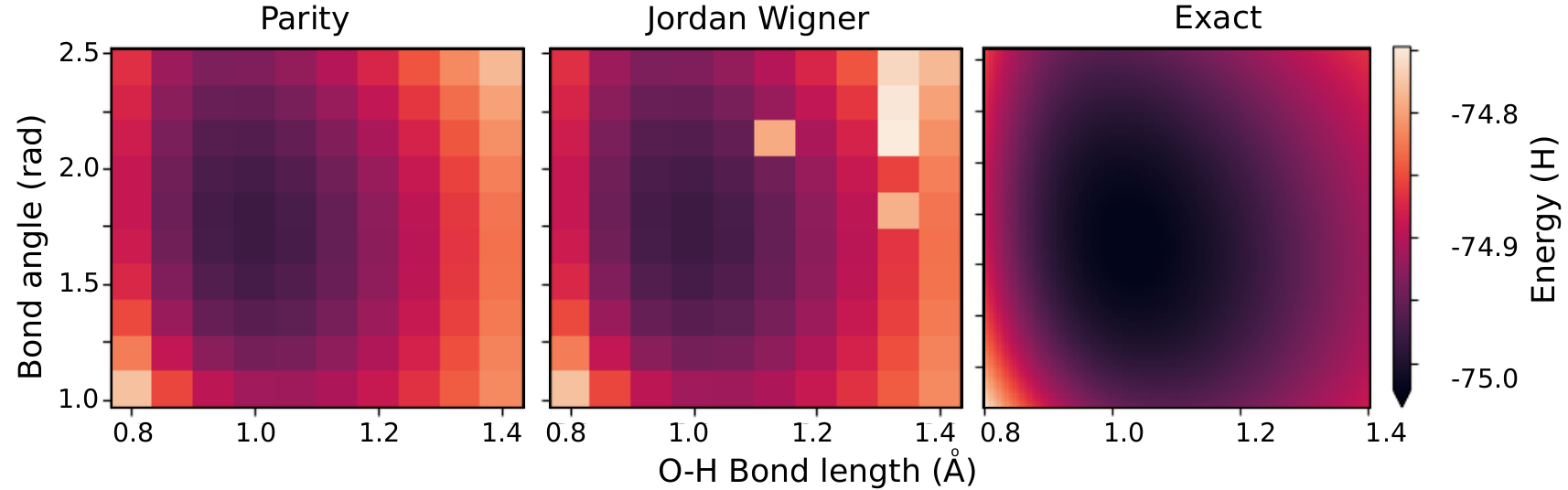}
	\caption{Ground state energy surface of H$_2$O. (left) Clifford pre-optimisation with Parity mapping  (mid) Clifford pre-optimisation with Jordan-Wigner mapping, and (right) exact diagonalization. There is a noticeable bias towards shorter distances/large bond angles where the pre-optimisation gets closer to the ground state. The patches in the Jordan-Wigner plot suggest the optimisation landscape under this transform is more difficult for the simulated annealing. The numerical results show that the energy obtained from Clifford Pre-optimisation can find a good initialization for the ground state energy.}
	\label{fig:H2O}
\end{figure*}

\subsection{Quantum chemistry}

Fig.~\ref{fig:H8} shows the effect of ansatz type, ansatz depth, and Fermion transform for a chain of eight hydrogen atoms. The H$_8$ chain is parameterized by the dimerisation length (see inset in Fig.~\ref{fig:H8}), and the plots show the best expectation value found over $10^4$ iterations of simulated annealing of the Clifford circuits (see appendix~\ref{section: numericals}). The real ansatz outperforms the Trotterized ansatz at depth $D=2$, indicating the Trotterized ansatz struggles to parameterize Clifford points close to the ground state despite having the same entanglement structure. In contrast, the Trotterized ansatz performs better at depth $D=5$ as it allows more flexible (complex valued) stabilizer states. Interestingly, although the depth $D=5$ real-valued ansatz can parameterise a super-set of the $D=2$ stabilizer states, it performs significantly worse. We attribute this to ansatz over-parameterization as the circuit expressibility and parameter dimension are both significantly increased. Correspondingly the $D=2$ Trotterized ansatz is underparameterized, as the more complex (and more difficult to optimize) $D=5$ circuit seems to find a good solution. There is also a noticeable difference between the Fermion transforms: each transform results gives a different set of weights and Pauli strings, and therefore the ground state has different coefficients in the logical basis. This places the ground state at different parts of the Hilbert space, and therefore has different projections onto the set of circuit-accessible stabilizer states. While here it looks like the Bravyi-Kitaev slightly underperforms compared to the Parity and Jordan-Wigner maps, we have no current understanding as to if this is expected in general.

Next, we consider the ten qubits H$_2$O Hamiltonian with varying O-H bond length and H-O-H bond angle, Fig.~\ref{fig:H2O}. Again we note the parity transform outperforms the Jordan-Wigner transform (Fig.~\ref{fig:H2O}). In both transformations, there is a noticeable bias in the minimum towards shorter bond lengths, and larger bond angles. That is, Clifford pre-optimisation finds a more accurate energy solution near shorter bond lengths and larger bond angles. Since quantum chemistry Hamiltonians have far more Pauli strings (typically $\mathcal{O}(poly(n))$) compared to solid-state ($\mathcal{O}(n)$) or binary optimisation ($\mathcal{O}(n^2)$), we expect them to benefit the most from circuit-agnostic Clifford pre-optimisation, due to the higher probability of finding non-zero expectation value stabilizer states.

\subsection{TFIM}
\begin{figure*}
	\centering
	\includegraphics[width=0.8\textwidth]{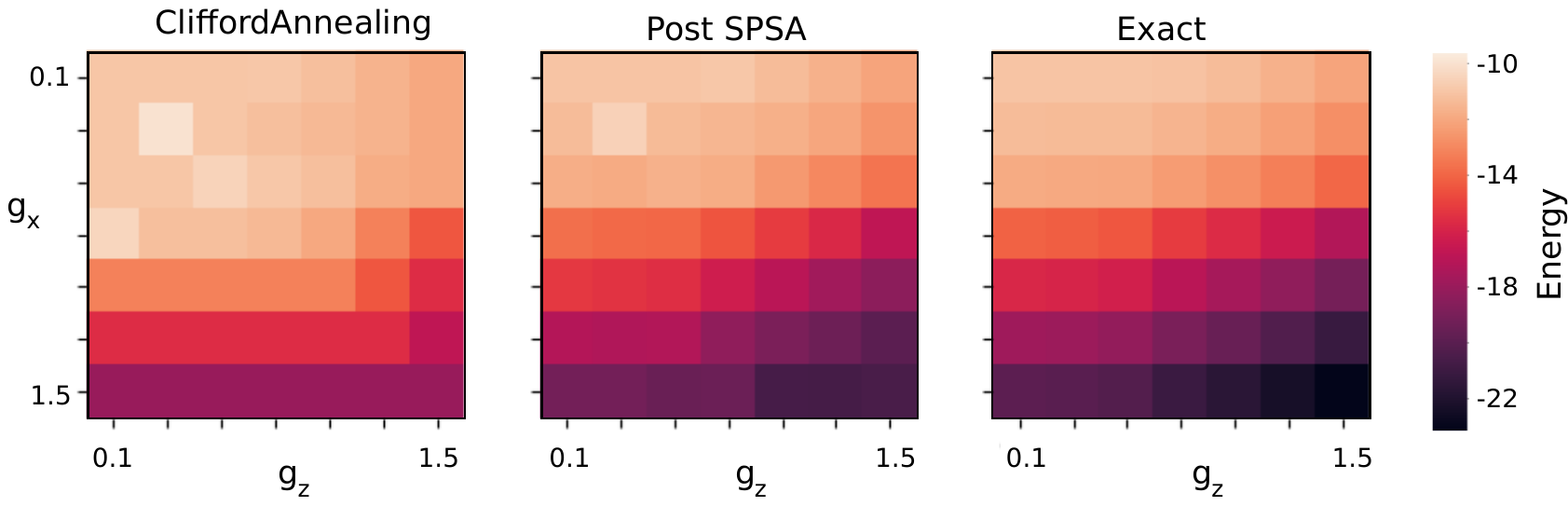}
	\caption{Eight-qubit transverse field Ising model [Eq.~\eqref{eq:H}] energy as a function of dimensionless transverse ($g_x$) and longitudinal ($g_z$) fields with $J=1$. Plots show the lowest energy seen over the first $10^4$ annealed stabilizer states (left), lowest energy after two hundred further rounds of SPSA starting from the best initial point (middle), and solution from exact diagonalization (right). The Clifford results (left) show that as $g_x$ becomes larger than $g_z$, the ansatz seems to find the same energy (independent of $g_z$). This suggests that the stabilizer state is preferentially finding $X_i$ eigenstates over $Z_i$, or $Z_iZ_{i+1}$ eigenstates (a similar effect is seen for large $g_z$ and $g_x\leq 0.5$). 
	}
	\label{fig:PhaseDiagram}
\end{figure*}

Here we benchmark the Clifford pre-optimisation with an eight-qubit TFIM Hamiltonian, with varying $g_x/J$ and $g_z/J$ fields (see Eq.~\eqref{eq:H} with $w_{ij} = J\delta_{i,i+1}$). Fig.~\ref{fig:PhaseDiagram} shows the Clifford search finds low energy states for low fields, but begins struggling for higher fields. We also plot the solution after two hundred rounds of continuous parameter SPSA descent to resolve the effects of pre-optimisation and circuit expressibility. For $g_x = g_z \approx 0.9$, the ground state exhibits long-range entanglement; however, since the continuous parameter gradient decent (using Simultaneous Perturbation Stochastic Approximation --- SPSA) quickly converges within two hundred iterations, this suggests the ansatz is sufficiently expressive for the required entanglement, but a non-trivial stabilizer solution isn't found.

For $g_x \geq g_z$, the Clifford ansatz finds the $X_i$ stabilizer states (i.e. the eigenstate of each of the Pauli $X_i$ operators), independent of the $g_z$ field. This can be understood by noting that $Z_i$ generates $Z_iZ_{i+1}$ terms in the stabilizer group (i.e. all $Z_iZ_{i+1}$ are products of the single $Z_i$ operators), thereby giving correlated eigenvalues as they all commute. These correlations preclude a low energy solution: the low energy (i.e. -1 eigenstates) states of $Z_i$ and $Z_{i+1}$, result in a positive energy contribution for $Z_iZ_{i+1}$ (i.e. is a +1 eigenstate), while no such restriction applies for $X_i$ and $X_{i+1}$ stabilizers. Likewise, all low energy (i.e. odd parity) stabilizer states of $Z_iZ_{i+1}$ will give identically zero energy for $X_i$, $X_{i+1}$ terms, and zero energy for $Z_i + Z_{i+1}$ terms\footnote{A low energy $\mean{Z_iZ_j}$ stabilizer state is either an odd parity logical state (in $i,j$) in which case  $\mean{Z_i} =\pm 1 = -\mean{Z_j}$, or a Bell-like state (in $i,j$) in which case $\mean{Z_i} = \mean{Z_j} = 0$ }. For large $g_x$, the Clifford pre-optimisation simply finds the ground state of the $J=g_z=0$ Hamiltonian, and there appears to be little benefit in running it. This emphasises that Clifford pre-optimisation can be parameter-specific and not just problem-specific.   

Following this logic, we could expect that when the Hamiltonian contains many non-commuting equally weighted Pauli strings, the lowest energy Clifford state may not give a good ground state estimate. Nonetheless, in some cases (e.g. $g_z \approx g_x$) it can give a substantially better initial condition, thereby allowing a continuous parameter VQE optimisation to convergence faster.

\subsection{MAXCUT Problem}
Finally, we consider Clifford pre-optimisation for the MAXCUT problem. The MAXCUT task is qualitatively different from quantum chemistry and TFIM tasks, as all of the $\mathcal{O}(n^2)$ Pauli Strings commute with each other. Furthermore, it is an example of a fundamentally classical problem of unconstrained quadratic binary optimisation.

When the ansatz circuit is not expressive enough to include the ideal stabilizer state (i.e. the solution), one might expect the Clifford search to be unsuccessful in finding a low expectation value initialisation. Interestingly, this is also the case for the problem-specific (QAOA) ansatz circuits. The QAOA ansatz can be readily tuned to a Clifford circuit (by setting all parameters to be multiples of $\pi/2$). However, due to the constrained nature of the circuit, the stabilizer search is not a useful initialisation strategy. For binary optimisation problems the target state is a logical basis state, and while there are guarantees that the optimal circuit gets close to the target basis states~\cite{QAOA}, the circuit cannot exactly parameterize a good stabilizer state. We note that since the target state is a logical basis state, the optimal Clifford ansatz is constructed only from a single layer of single-qubit X-gates --- in which case the Clifford search trivially becomes identical to an implementation of classical simulated annealing.

\section{Scaling}
\label{section:scaling}

\begin{figure}
	\centering
	\includegraphics[width=\columnwidth]{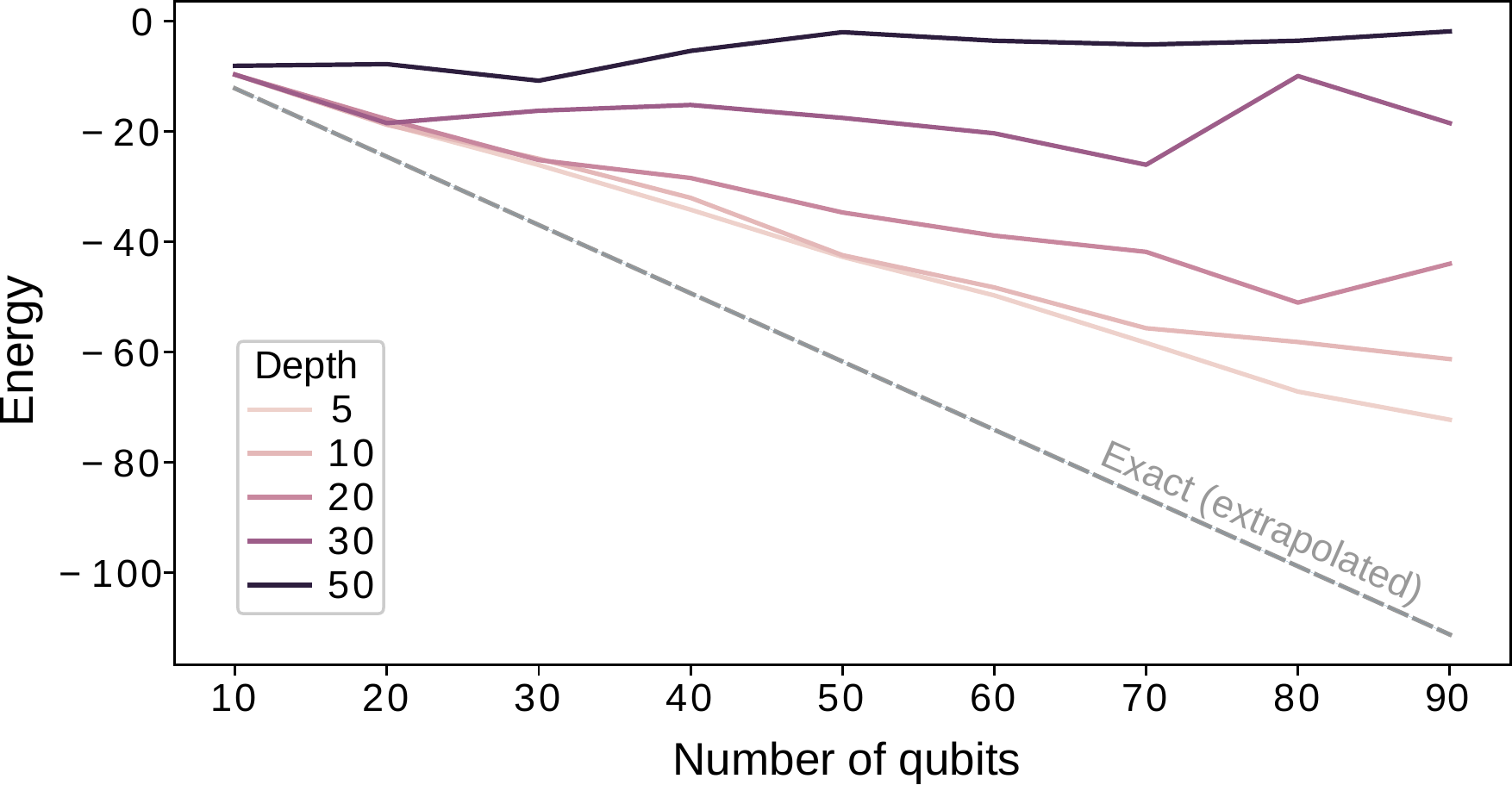}
	\caption{Minimum pre-optimised energy for the TFIM problem ($g_z=g_x=0.85J$), as the number of qubits increases, but keeping the Clifford annealing time fixed to $10^4$ iterations. The dashed line shows the (linearly extrapolated) ground state. The relationship between the depth (different colours) and number of qubits suggest the method is not arbitrarily salable for the TFIM, but for shallower circuits and several tens of qubits, Clifford pre-optimisation can find a good initial condition.}
	\label{fig:tfim_scale}
\end{figure}

Here we analytically argue that a Clifford based initialisation is not arbitrarily scalable, but due to practical considerations, may still be useful for VQE tasks involving several tens of  qubits. In agreement with the numerical results above, we show that the method is particularly attractive for quantum chemistry problems, compared to binary optimisation or solid-state problems. 

The ansatz circuit evaluated at the Clifford parameter points results in a stabilizer state. The utility of the pre-optimisation depends on how many stabilizer states give a non-zero expectation value of the cost function operator $H$. For an $n$ qubit cost function, this can be estimated given the Pauli string decomposition, $H = \sum_i c_i P_i$, for Pauli strings $P_i$ and coefficients $c_i$. Stabilizer states ($\ket{s}$) either return zero expectation value $\Sij{s}{P}{s} = 0$ for a Pauli string $P$, or are eigenstates $P\ket{s} = \pm \ket{s}$, in which case $P$ is said to stabilize $\ket{s}$.

Following Ref.~\cite{Aaronson_2004}, the number $N_\mathrm{P}$ of stabilizer states which can be stabilized by a single $n$ qubit Pauli string is $N_\mathrm{P} = S(n - 1)$, with 
\begin{eqnarray}
	S(n) = 2^n \prod^n_{k = 1}(2^k + 1)
\end{eqnarray}
while the total number of stabilizer states is given by $N_\mathrm{T} = S(n)$. $N_\mathrm{P}$ is exponentially smaller than $N_\mathrm{T}$. Hence, for a Hamiltonian which is decomposed into $poly(n)$ Pauli strings, the probability of getting a non-zero expectation value becomes vanishing small asymptotically. Therefore in the most general case, to keep up the scaling of the stabilizer state space, the number of Pauli Strings in the Hamiltonian must also grow exponentially. Furthermore, anti-commuting Pauli strings do not share any stabilizers, and two commuting strings share a set of $N_\mathrm{C} = S(n - 2)$ common stabilizers. Hence, finding a single stabilizer state that is stabilized by two commuting Pauli strings also becomes vanishing small asymptotically. 

Combing the above counting arguments, the number of stabilizer states that give a non-zero expectation value to the Hamiltonian ($N_\mathrm{H}$) has an upper bound, $N_\mathrm{H} < N_\mathrm{P} M$, where $M$ is the number of terms in the Pauli decomposition of $H$. If $M$ scales as $poly(n)$, then the Hamiltonian cannot escape the fate of having an exponentially shrinking size of stabilizer states, giving an effective barren plateau for Clifford points. This exponentially growing zero-eigenvalue stabilizer state will become the main obstacle to scaling up the Clifford pre-optimisation, and is equivalent to the barren plateau in the continuous parameter case. Nevertheless, stabilizer state pre-optimisation is useful for device bench-marking, and possible (non-scalable) small-scale demonstration of quantum advantage, and selecting useful ansatz circuits.

For less-than-linear depth circuits, the number of accessible stabilizer states is far smaller than the (exponential in $n^2$) number of all stabilizer states, $N_T$ (this is also the case for any depth circuits with a restricted number of parameterized gates). This smaller set of circuit-constrained set of states could allow a higher-than-random chance of a given state being stabilized by Pauli strings in the Hamiltonian. This has the potential for limited scalability of classical preoptimisatoin, perhaps into the quantum advantage regime, see Fig.~\ref{fig:tfim_scale}. Moreover, classically finding a good initial state, or suitable ansatz circuit will allow increased throughput of quantum devices. 

Section \ref{section: results} showed the quantum chemistry problems (with their many groups of non-commuting Pauli strings), benefited more from the Clifford search, when compared to the TFIM (with only two groups of non-commuting Pauli strings), or MAXCUT (where all Pauli strings commute). Scaling arguments notwithstanding, the more non-commuting Pauli strings in the Hamiltonian, the more useful a stabilizer state search. We reiterate that quantum chemistry problems fit this paradigm, suggesting that quantum chemistry problems, in contrast to binary optimisation and solid-state models, are particularly well suited to stabilizer pre-optimisation in the ISQ regime.

\section{Conclusion}

In this paper, we presented Clifford pre-optimisation; a classical tool that reduces the load on the quantum part for the VQE. This algorithm optimizes for stabilizer states accessible from a given quantum circuit (with the aid of Clifford gates) to minimize a cost function. Many stabilizer states are accessible by a parameterized circuit, and these uniformly distributed, problem independent states, make them ideal for comparing the cost functions before a quantum device is used for full continuous parameter optimization. The total number of stabilizers scales faster-than-exponentially with the number of qubits (exponential with $n^2$), but exponentially with the number of ansatz parameters, thereby highlighting the importance of the ansatz circuit. Furthermore, the number of stabilizer states that are Pauli string eigenvalues also scales exponentially in $n^2$ (although it is exponentially small compared to the total numbers of stabilizer states), but a given restricted depth ansatz circuit does not uniformly sample all states, potentially allowing more useful states to be found. 

We presented numerical studies using Clifford circuit annealing as the pre-optimisation method for quantum chemistry, TFIM, and QAOA problems. We unraveled the effects of circuit choice, Hamiltonian structure (global, local, and the number of commuting/non-commuting Pauli terms), and Fermion transform on the pre-optimisations performance. We identified that the Clifford search finds a single stabilizer of a (commuting) set of Pauli strings in the cost function, and this is not necessarily the maximally commuting set, Fig.~\ref{fig:PhaseDiagram}. 

Our results show that Clifford pre-optimisation provides a powerful new method to lower the cost of required quantum processing time but pushing the classical limit of the VQE. The method is useful for both device bench marking, and likely (albeit not provably) useful for ISQ quantum advantage, in particular for quantum chemistry problems, Fig.~\ref{fig:H8} and \ref{fig:H2O}. However, we noted that for many qubits and deep circuits, the counting arguments of Sec.~\ref{section:scaling}, show the eventual onset of a barren plateau, and limit the arbitrary scalability of the method, even when individual Clifford circuits themselves are efficiently evaluated.

\section{Acknowledgements}
We thank Tobias Hung for his helpful insights during this work, and David Amaro for constructive comments on the manuscript. KEK and MSK, acknowledge funding from the Samsung GRC grant, KIST grant and EPSRC Quantum computing and simulation hub grant. MHC and ACM acknowledge funding from the Fraunhofer Institute. Source code available on request.

\bibliography{Clifford.bib}

\begin{thebibliography}{37}%
\makeatletter
\providecommand \@ifxundefined [1]{%
 \@ifx{#1\undefined}
}%
\providecommand \@ifnum [1]{%
 \ifnum #1\expandafter \@firstoftwo
 \else \expandafter \@secondoftwo
 \fi
}%
\providecommand \@ifx [1]{%
 \ifx #1\expandafter \@firstoftwo
 \else \expandafter \@secondoftwo
 \fi
}%
\providecommand \natexlab [1]{#1}%
\providecommand \enquote  [1]{``#1''}%
\providecommand \bibnamefont  [1]{#1}%
\providecommand \bibfnamefont [1]{#1}%
\providecommand \citenamefont [1]{#1}%
\providecommand \href@noop [0]{\@secondoftwo}%
\providecommand \href [0]{\begingroup \@sanitize@url \@href}%
\providecommand \@href[1]{\@@startlink{#1}\@@href}%
\providecommand \@@href[1]{\endgroup#1\@@endlink}%
\providecommand \@sanitize@url [0]{\catcode `\\12\catcode `\$12\catcode
  `\&12\catcode `\#12\catcode `\^12\catcode `\_12\catcode `\%12\relax}%
\providecommand \@@startlink[1]{}%
\providecommand \@@endlink[0]{}%
\providecommand \url  [0]{\begingroup\@sanitize@url \@url }%
\providecommand \@url [1]{\endgroup\@href {#1}{\urlprefix }}%
\providecommand \urlprefix  [0]{URL }%
\providecommand \Eprint [0]{\href }%
\providecommand \doibase [0]{https://doi.org/}%
\providecommand \selectlanguage [0]{\@gobble}%
\providecommand \bibinfo  [0]{\@secondoftwo}%
\providecommand \bibfield  [0]{\@secondoftwo}%
\providecommand \translation [1]{[#1]}%
\providecommand \BibitemOpen [0]{}%
\providecommand \bibitemStop [0]{}%
\providecommand \bibitemNoStop [0]{.\EOS\space}%
\providecommand \EOS [0]{\spacefactor3000\relax}%
\providecommand \BibitemShut  [1]{\csname bibitem#1\endcsname}%
\let\auto@bib@innerbib\@empty
\bibitem [{\citenamefont {Kitaev}(1995)}]{phase_est}%
  \BibitemOpen
  \bibfield  {author} {\bibinfo {author} {\bibfnamefont {A.~Y.}\ \bibnamefont
  {Kitaev}},\ }\href {https://doi.org/10.48550/ARXIV.QUANT-PH/9511026}
  {\bibinfo {title} {Quantum measurements and the abelian stabilizer problem}}
  (\bibinfo {year} {1995})\BibitemShut {NoStop}%
\bibitem [{\citenamefont {Shor}(1994)}]{Shor}%
  \BibitemOpen
  \bibfield  {author} {\bibinfo {author} {\bibfnamefont {P.}~\bibnamefont
  {Shor}},\ }\bibfield  {title} {\bibinfo {title} {Algorithms for quantum
  computation: discrete logarithms and factoring},\ }in\ \href
  {https://doi.org/10.1109/SFCS.1994.365700} {\emph {\bibinfo {booktitle}
  {Proceedings 35th Annual Symposium on Foundations of Computer Science}}}\
  (\bibinfo {year} {1994})\ pp.\ \bibinfo {pages} {124--134}\BibitemShut
  {NoStop}%
\bibitem [{\citenamefont {Deutsch}\ and\ \citenamefont
  {Jozsa}(1992)}]{Deutsch}%
  \BibitemOpen
  \bibfield  {author} {\bibinfo {author} {\bibfnamefont {D.}~\bibnamefont
  {Deutsch}}\ and\ \bibinfo {author} {\bibfnamefont {R.}~\bibnamefont
  {Jozsa}},\ }\bibfield  {title} {\bibinfo {title} {Rapid solution of problems
  by quantum computation},\ }\href {https://doi.org/10.1098/rspa.1992.0167}
  {\bibfield  {journal} {\bibinfo  {journal} {Proceedings of the Royal Society
  A}\ }\textbf {\bibinfo {volume} {439}},\ \bibinfo {pages} {553–558}
  (\bibinfo {year} {1992})}\BibitemShut {NoStop}%
\bibitem [{\citenamefont {Peruzzo}\ \emph {et~al.}(2014)\citenamefont
  {Peruzzo}, \citenamefont {McClean}, \citenamefont {Shadbolt}, \citenamefont
  {Yung}, \citenamefont {Zhou},\ and\ \citenamefont {Love}}]{VQE_1}%
  \BibitemOpen
  \bibfield  {author} {\bibinfo {author} {\bibfnamefont {A.}~\bibnamefont
  {Peruzzo}}, \bibinfo {author} {\bibfnamefont {J.}~\bibnamefont {McClean}},
  \bibinfo {author} {\bibfnamefont {P.}~\bibnamefont {Shadbolt}}, \bibinfo
  {author} {\bibfnamefont {M.~H.}\ \bibnamefont {Yung}}, \bibinfo {author}
  {\bibfnamefont {X.~Q.}\ \bibnamefont {Zhou}},\ and\ \bibinfo {author}
  {\bibfnamefont {P.~J.}\ \bibnamefont {Love}},\ }\bibfield  {title} {\bibinfo
  {title} {A variational eigenvalue solver on a photonic quantum processor},\
  }\href {https://doi.org/10.1038/ncomms5213} {\bibfield  {journal} {\bibinfo
  {journal} {Nature Communications}\ }\textbf {\bibinfo {volume} {5}},\
  \bibinfo {pages} {4213} (\bibinfo {year} {2014})}\BibitemShut {NoStop}%
\bibitem [{\citenamefont {McClean}\ \emph {et~al.}(2016)\citenamefont
  {McClean}, \citenamefont {Romero}, \citenamefont {Babbush},\ and\
  \citenamefont {Aspuru-Guzik}}]{VQE_2}%
  \BibitemOpen
  \bibfield  {author} {\bibinfo {author} {\bibfnamefont {J.~R.}\ \bibnamefont
  {McClean}}, \bibinfo {author} {\bibfnamefont {J.}~\bibnamefont {Romero}},
  \bibinfo {author} {\bibfnamefont {R.}~\bibnamefont {Babbush}},\ and\ \bibinfo
  {author} {\bibfnamefont {A.}~\bibnamefont {Aspuru-Guzik}},\ }\bibfield
  {title} {\bibinfo {title} {The theory of variational hybrid quantum-classical
  algorithms},\ }\href {https://doi.org/10.1088/1367-2630/18/2/023023}
  {\bibfield  {journal} {\bibinfo  {journal} {New Journal of Physics}\ }\textbf
  {\bibinfo {volume} {18}},\ \bibinfo {pages} {023023} (\bibinfo {year}
  {2016})}\BibitemShut {NoStop}%
\bibitem [{\citenamefont {Romero}\ \emph {et~al.}(2018)\citenamefont {Romero},
  \citenamefont {Babbush}, \citenamefont {McClean}, \citenamefont {Hempel},
  \citenamefont {Love},\ and\ \citenamefont {Aspuru-Guzik}}]{VQE_3}%
  \BibitemOpen
  \bibfield  {author} {\bibinfo {author} {\bibfnamefont {J.}~\bibnamefont
  {Romero}}, \bibinfo {author} {\bibfnamefont {R.}~\bibnamefont {Babbush}},
  \bibinfo {author} {\bibfnamefont {J.~R.}\ \bibnamefont {McClean}}, \bibinfo
  {author} {\bibfnamefont {C.}~\bibnamefont {Hempel}}, \bibinfo {author}
  {\bibfnamefont {P.~J.}\ \bibnamefont {Love}},\ and\ \bibinfo {author}
  {\bibfnamefont {A.}~\bibnamefont {Aspuru-Guzik}},\ }\bibfield  {title}
  {\bibinfo {title} {Strategies for quantum computing molecular energies using
  the unitary coupled cluster ansatz},\ }\href
  {https://doi.org/10.1088/2058-9565/aad3e4} {\bibfield  {journal} {\bibinfo
  {journal} {Quantum Science and Technology}\ }\textbf {\bibinfo {volume}
  {4}},\ \bibinfo {pages} {014008} (\bibinfo {year} {2018})}\BibitemShut
  {NoStop}%
\bibitem [{\citenamefont {Fran{\c{c}}a}\ and\ \citenamefont
  {Garc{\'{\i}}a-Patr{\'{o}}n}(2021)}]{Noise_limitation}%
  \BibitemOpen
  \bibfield  {author} {\bibinfo {author} {\bibfnamefont {D.~S.}\ \bibnamefont
  {Fran{\c{c}}a}}\ and\ \bibinfo {author} {\bibfnamefont {R.}~\bibnamefont
  {Garc{\'{\i}}a-Patr{\'{o}}n}},\ }\bibfield  {title} {\bibinfo {title}
  {Limitations of optimization algorithms on noisy quantum devices},\ }\href
  {https://doi.org/10.1038/s41567-021-01356-3} {\bibfield  {journal} {\bibinfo
  {journal} {Nature Physics}\ }\textbf {\bibinfo {volume} {17}},\ \bibinfo
  {pages} {1221} (\bibinfo {year} {2021})}\BibitemShut {NoStop}%
\bibitem [{\citenamefont {Farhi}\ \emph {et~al.}(2014)\citenamefont {Farhi},
  \citenamefont {Goldstone},\ and\ \citenamefont {Gutmann}}]{QAOA}%
  \BibitemOpen
  \bibfield  {author} {\bibinfo {author} {\bibfnamefont {E.}~\bibnamefont
  {Farhi}}, \bibinfo {author} {\bibfnamefont {J.}~\bibnamefont {Goldstone}},\
  and\ \bibinfo {author} {\bibfnamefont {S.}~\bibnamefont {Gutmann}},\ }\href
  {https://doi.org/10.48550/ARXIV.1411.4028} {\bibinfo {title} {A quantum
  approximate optimization algorithm}} (\bibinfo {year} {2014})\BibitemShut
  {NoStop}%
\bibitem [{\citenamefont {Amaro}\ \emph {et~al.}(2022)\citenamefont {Amaro},
  \citenamefont {Modica}, \citenamefont {Rosenkranz}, \citenamefont
  {Fiorentini}, \citenamefont {Benedetti},\ and\ \citenamefont
  {Lubasch}}]{amaro_filtering_2022}%
  \BibitemOpen
  \bibfield  {author} {\bibinfo {author} {\bibfnamefont {D.}~\bibnamefont
  {Amaro}}, \bibinfo {author} {\bibfnamefont {C.}~\bibnamefont {Modica}},
  \bibinfo {author} {\bibfnamefont {M.}~\bibnamefont {Rosenkranz}}, \bibinfo
  {author} {\bibfnamefont {M.}~\bibnamefont {Fiorentini}}, \bibinfo {author}
  {\bibfnamefont {M.}~\bibnamefont {Benedetti}},\ and\ \bibinfo {author}
  {\bibfnamefont {M.}~\bibnamefont {Lubasch}},\ }\bibfield  {title} {\bibinfo
  {title} {Filtering variational quantum algorithms for combinatorial
  optimization},\ }\href {https://doi.org/10.1088/2058-9565/ac3e54} {\bibfield
  {journal} {\bibinfo  {journal} {Quantum Sci. Technol.}\ }\textbf {\bibinfo
  {volume} {7}},\ \bibinfo {pages} {015021} (\bibinfo {year} {2022})},\
  \bibinfo {note} {arXiv:2106.10055 [quant-ph]}\BibitemShut {NoStop}%
\bibitem [{\citenamefont {Orús}\ \emph {et~al.}(2019)\citenamefont {Orús},
  \citenamefont {Mugel},\ and\ \citenamefont {Lizaso}}]{Profolio_Opt}%
  \BibitemOpen
  \bibfield  {author} {\bibinfo {author} {\bibfnamefont {R.}~\bibnamefont
  {Orús}}, \bibinfo {author} {\bibfnamefont {S.}~\bibnamefont {Mugel}},\ and\
  \bibinfo {author} {\bibfnamefont {E.}~\bibnamefont {Lizaso}},\ }\bibfield
  {title} {\bibinfo {title} {Quantum computing for finance: Overview and
  prospects},\ }\href
  {https://doi.org/https://doi.org/10.1016/j.revip.2019.100028} {\bibfield
  {journal} {\bibinfo  {journal} {Reviews in Physics}\ }\textbf {\bibinfo
  {volume} {4}},\ \bibinfo {pages} {100028} (\bibinfo {year}
  {2019})}\BibitemShut {NoStop}%
\bibitem [{\citenamefont {Braine}\ \emph {et~al.}(2021)\citenamefont {Braine},
  \citenamefont {Egger}, \citenamefont {Glick},\ and\ \citenamefont
  {Woerner}}]{financial_settlement}%
  \BibitemOpen
  \bibfield  {author} {\bibinfo {author} {\bibfnamefont {L.}~\bibnamefont
  {Braine}}, \bibinfo {author} {\bibfnamefont {D.~J.}\ \bibnamefont {Egger}},
  \bibinfo {author} {\bibfnamefont {J.}~\bibnamefont {Glick}},\ and\ \bibinfo
  {author} {\bibfnamefont {S.}~\bibnamefont {Woerner}},\ }\bibfield  {title}
  {\bibinfo {title} {Quantum algorithms for mixed binary optimization applied
  to transaction settlement},\ }\href
  {https://doi.org/10.1109/tqe.2021.3063635} {\bibfield  {journal} {\bibinfo
  {journal} {{IEEE} Transactions on Quantum Engineering}\ }\textbf {\bibinfo
  {volume} {2}},\ \bibinfo {pages} {1} (\bibinfo {year} {2021})}\BibitemShut
  {NoStop}%
\bibitem [{\citenamefont {Ma}\ \emph {et~al.}(2020)\citenamefont {Ma},
  \citenamefont {Govoni},\ and\ \citenamefont {Galli}}]{solid-state}%
  \BibitemOpen
  \bibfield  {author} {\bibinfo {author} {\bibfnamefont {H.}~\bibnamefont
  {Ma}}, \bibinfo {author} {\bibfnamefont {M.}~\bibnamefont {Govoni}},\ and\
  \bibinfo {author} {\bibfnamefont {G.}~\bibnamefont {Galli}},\ }\bibfield
  {title} {\bibinfo {title} {Quantum simulations of materials on near-term
  quantum computers},\ }\href {https://doi.org/10.1038/s41524-020-00353-z}
  {\bibfield  {journal} {\bibinfo  {journal} {npj Computational Materials}\
  }\textbf {\bibinfo {volume} {6}},\ \bibinfo {pages} {85} (\bibinfo {year}
  {2020})}\BibitemShut {NoStop}%
\bibitem [{\citenamefont {Consiglio}\ \emph {et~al.}(2021)\citenamefont
  {Consiglio}, \citenamefont {Chetcuti}, \citenamefont {Bravo-Prieto},
  \citenamefont {Ramos-Calderer}, \citenamefont {Minguzzi}, \citenamefont
  {Latorre}, \citenamefont {Amico},\ and\ \citenamefont
  {Apollaro}}]{SU(N)_fermions}%
  \BibitemOpen
  \bibfield  {author} {\bibinfo {author} {\bibfnamefont {M.}~\bibnamefont
  {Consiglio}}, \bibinfo {author} {\bibfnamefont {W.~J.}\ \bibnamefont
  {Chetcuti}}, \bibinfo {author} {\bibfnamefont {C.}~\bibnamefont
  {Bravo-Prieto}}, \bibinfo {author} {\bibfnamefont {S.}~\bibnamefont
  {Ramos-Calderer}}, \bibinfo {author} {\bibfnamefont {A.}~\bibnamefont
  {Minguzzi}}, \bibinfo {author} {\bibfnamefont {J.~I.}\ \bibnamefont
  {Latorre}}, \bibinfo {author} {\bibfnamefont {L.}~\bibnamefont {Amico}},\
  and\ \bibinfo {author} {\bibfnamefont {T.~J.~G.}\ \bibnamefont {Apollaro}},\
  }\href {https://doi.org/10.48550/ARXIV.2106.15552} {\bibinfo {title}
  {Variational quantum eigensolver for su($n$) fermions}} (\bibinfo {year}
  {2021})\BibitemShut {NoStop}%
\bibitem [{\citenamefont {Arute}\ \emph {et~al.}(2020)\citenamefont {Arute},
  \citenamefont {Arya}, \citenamefont {Babbush}, \citenamefont {Bacon},
  \citenamefont {Bardin}, \citenamefont {Barends}, \citenamefont {Boixo},
  \citenamefont {Broughton}, \citenamefont {Buckley}, \citenamefont {Buell},
  \citenamefont {Burkett}, \citenamefont {Bushnell}, \citenamefont {Chen},
  \citenamefont {Chen}, \citenamefont {Chiaro}, \citenamefont {Collins},
  \citenamefont {Courtney}, \citenamefont {Demura}, \citenamefont {Dunsworth},
  \citenamefont {Farhi}, \citenamefont {Fowler}, \citenamefont {Foxen},
  \citenamefont {Gidney}, \citenamefont {Giustina}, \citenamefont {Graff},
  \citenamefont {Habegger}, \citenamefont {Harrigan}, \citenamefont {Ho},
  \citenamefont {Hong}, \citenamefont {Huang}, \citenamefont {Huggins},
  \citenamefont {Ioffe}, \citenamefont {Isakov}, \citenamefont {Jeffrey},
  \citenamefont {Jiang}, \citenamefont {Jones}, \citenamefont {Kafri},
  \citenamefont {Kechedzhi}, \citenamefont {Kelly}, \citenamefont {Kim},
  \citenamefont {Klimov}, \citenamefont {Korotkov}, \citenamefont {Kostritsa},
  \citenamefont {Landhuis}, \citenamefont {Laptev}, \citenamefont {Lindmark},
  \citenamefont {Lucero}, \citenamefont {Martin}, \citenamefont {Martinis},
  \citenamefont {McClean}, \citenamefont {McEwen}, \citenamefont {Megrant},
  \citenamefont {Mi}, \citenamefont {Mohseni}, \citenamefont {Mruczkiewicz},
  \citenamefont {Mutus}, \citenamefont {Naaman}, \citenamefont {Neeley},
  \citenamefont {Neill}, \citenamefont {Neven}, \citenamefont {Niu},
  \citenamefont {O’Brien}, \citenamefont {Ostby}, \citenamefont {Petukhov},
  \citenamefont {Putterman}, \citenamefont {Quintana}, \citenamefont {Roushan},
  \citenamefont {Rubin}, \citenamefont {Sank}, \citenamefont {Satzinger},
  \citenamefont {Smelyanskiy}, \citenamefont {Strain}, \citenamefont {Sung},
  \citenamefont {Szalay}, \citenamefont {Takeshita}, \citenamefont
  {Vainsencher}, \citenamefont {White}, \citenamefont {Wiebe}, \citenamefont
  {Yao}, \citenamefont {Yeh},\ and\ \citenamefont {Zalcman}}]{Hartree-Fock}%
  \BibitemOpen
  \bibfield  {author} {\bibinfo {author} {\bibfnamefont {F.}~\bibnamefont
  {Arute}}, \bibinfo {author} {\bibfnamefont {K.}~\bibnamefont {Arya}},
  \bibinfo {author} {\bibfnamefont {R.}~\bibnamefont {Babbush}}, \bibinfo
  {author} {\bibfnamefont {D.}~\bibnamefont {Bacon}}, \bibinfo {author}
  {\bibfnamefont {J.~C.}\ \bibnamefont {Bardin}}, \bibinfo {author}
  {\bibfnamefont {R.}~\bibnamefont {Barends}}, \bibinfo {author} {\bibfnamefont
  {S.}~\bibnamefont {Boixo}}, \bibinfo {author} {\bibfnamefont
  {M.}~\bibnamefont {Broughton}}, \bibinfo {author} {\bibfnamefont {B.~B.}\
  \bibnamefont {Buckley}}, \bibinfo {author} {\bibfnamefont {D.~A.}\
  \bibnamefont {Buell}}, \bibinfo {author} {\bibfnamefont {B.}~\bibnamefont
  {Burkett}}, \bibinfo {author} {\bibfnamefont {N.}~\bibnamefont {Bushnell}},
  \bibinfo {author} {\bibfnamefont {Y.}~\bibnamefont {Chen}}, \bibinfo {author}
  {\bibfnamefont {Z.}~\bibnamefont {Chen}}, \bibinfo {author} {\bibfnamefont
  {B.}~\bibnamefont {Chiaro}}, \bibinfo {author} {\bibfnamefont
  {R.}~\bibnamefont {Collins}}, \bibinfo {author} {\bibfnamefont
  {W.}~\bibnamefont {Courtney}}, \bibinfo {author} {\bibfnamefont
  {S.}~\bibnamefont {Demura}}, \bibinfo {author} {\bibfnamefont
  {A.}~\bibnamefont {Dunsworth}}, \bibinfo {author} {\bibfnamefont
  {E.}~\bibnamefont {Farhi}}, \bibinfo {author} {\bibfnamefont
  {A.}~\bibnamefont {Fowler}}, \bibinfo {author} {\bibfnamefont
  {B.}~\bibnamefont {Foxen}}, \bibinfo {author} {\bibfnamefont
  {C.}~\bibnamefont {Gidney}}, \bibinfo {author} {\bibfnamefont
  {M.}~\bibnamefont {Giustina}}, \bibinfo {author} {\bibfnamefont
  {R.}~\bibnamefont {Graff}}, \bibinfo {author} {\bibfnamefont
  {S.}~\bibnamefont {Habegger}}, \bibinfo {author} {\bibfnamefont {M.~P.}\
  \bibnamefont {Harrigan}}, \bibinfo {author} {\bibfnamefont {A.}~\bibnamefont
  {Ho}}, \bibinfo {author} {\bibfnamefont {S.}~\bibnamefont {Hong}}, \bibinfo
  {author} {\bibfnamefont {T.}~\bibnamefont {Huang}}, \bibinfo {author}
  {\bibfnamefont {W.~J.}\ \bibnamefont {Huggins}}, \bibinfo {author}
  {\bibfnamefont {L.}~\bibnamefont {Ioffe}}, \bibinfo {author} {\bibfnamefont
  {S.~V.}\ \bibnamefont {Isakov}}, \bibinfo {author} {\bibfnamefont
  {E.}~\bibnamefont {Jeffrey}}, \bibinfo {author} {\bibfnamefont
  {Z.}~\bibnamefont {Jiang}}, \bibinfo {author} {\bibfnamefont
  {C.}~\bibnamefont {Jones}}, \bibinfo {author} {\bibfnamefont
  {D.}~\bibnamefont {Kafri}}, \bibinfo {author} {\bibfnamefont
  {K.}~\bibnamefont {Kechedzhi}}, \bibinfo {author} {\bibfnamefont
  {J.}~\bibnamefont {Kelly}}, \bibinfo {author} {\bibfnamefont
  {S.}~\bibnamefont {Kim}}, \bibinfo {author} {\bibfnamefont {P.~V.}\
  \bibnamefont {Klimov}}, \bibinfo {author} {\bibfnamefont {A.}~\bibnamefont
  {Korotkov}}, \bibinfo {author} {\bibfnamefont {F.}~\bibnamefont {Kostritsa}},
  \bibinfo {author} {\bibfnamefont {D.}~\bibnamefont {Landhuis}}, \bibinfo
  {author} {\bibfnamefont {P.}~\bibnamefont {Laptev}}, \bibinfo {author}
  {\bibfnamefont {M.}~\bibnamefont {Lindmark}}, \bibinfo {author}
  {\bibfnamefont {E.}~\bibnamefont {Lucero}}, \bibinfo {author} {\bibfnamefont
  {O.}~\bibnamefont {Martin}}, \bibinfo {author} {\bibfnamefont {J.~M.}\
  \bibnamefont {Martinis}}, \bibinfo {author} {\bibfnamefont {J.~R.}\
  \bibnamefont {McClean}}, \bibinfo {author} {\bibfnamefont {M.}~\bibnamefont
  {McEwen}}, \bibinfo {author} {\bibfnamefont {A.}~\bibnamefont {Megrant}},
  \bibinfo {author} {\bibfnamefont {X.}~\bibnamefont {Mi}}, \bibinfo {author}
  {\bibfnamefont {M.}~\bibnamefont {Mohseni}}, \bibinfo {author} {\bibfnamefont
  {W.}~\bibnamefont {Mruczkiewicz}}, \bibinfo {author} {\bibfnamefont
  {J.}~\bibnamefont {Mutus}}, \bibinfo {author} {\bibfnamefont
  {O.}~\bibnamefont {Naaman}}, \bibinfo {author} {\bibfnamefont
  {M.}~\bibnamefont {Neeley}}, \bibinfo {author} {\bibfnamefont
  {C.}~\bibnamefont {Neill}}, \bibinfo {author} {\bibfnamefont
  {H.}~\bibnamefont {Neven}}, \bibinfo {author} {\bibfnamefont {M.~Y.}\
  \bibnamefont {Niu}}, \bibinfo {author} {\bibfnamefont {T.~E.}\ \bibnamefont
  {O’Brien}}, \bibinfo {author} {\bibfnamefont {E.}~\bibnamefont {Ostby}},
  \bibinfo {author} {\bibfnamefont {A.}~\bibnamefont {Petukhov}}, \bibinfo
  {author} {\bibfnamefont {H.}~\bibnamefont {Putterman}}, \bibinfo {author}
  {\bibfnamefont {C.}~\bibnamefont {Quintana}}, \bibinfo {author}
  {\bibfnamefont {P.}~\bibnamefont {Roushan}}, \bibinfo {author} {\bibfnamefont
  {N.~C.}\ \bibnamefont {Rubin}}, \bibinfo {author} {\bibfnamefont
  {D.}~\bibnamefont {Sank}}, \bibinfo {author} {\bibfnamefont {K.~J.}\
  \bibnamefont {Satzinger}}, \bibinfo {author} {\bibfnamefont {V.}~\bibnamefont
  {Smelyanskiy}}, \bibinfo {author} {\bibfnamefont {D.}~\bibnamefont {Strain}},
  \bibinfo {author} {\bibfnamefont {K.~J.}\ \bibnamefont {Sung}}, \bibinfo
  {author} {\bibfnamefont {M.}~\bibnamefont {Szalay}}, \bibinfo {author}
  {\bibfnamefont {T.~Y.}\ \bibnamefont {Takeshita}}, \bibinfo {author}
  {\bibfnamefont {A.}~\bibnamefont {Vainsencher}}, \bibinfo {author}
  {\bibfnamefont {T.}~\bibnamefont {White}}, \bibinfo {author} {\bibfnamefont
  {N.}~\bibnamefont {Wiebe}}, \bibinfo {author} {\bibfnamefont {Z.~J.}\
  \bibnamefont {Yao}}, \bibinfo {author} {\bibfnamefont {P.}~\bibnamefont
  {Yeh}},\ and\ \bibinfo {author} {\bibfnamefont {A.}~\bibnamefont {Zalcman}},\
  }\bibfield  {title} {\bibinfo {title} {Hartree-fock on a superconducting
  qubit quantum computer},\ }\href {https://doi.org/10.1126/science.abb9811}
  {\bibfield  {journal} {\bibinfo  {journal} {Science}\ }\textbf {\bibinfo
  {volume} {369}},\ \bibinfo {pages} {1084} (\bibinfo {year} {2020})},\ \Eprint
  {https://arxiv.org/abs/https://www.science.org/doi/pdf/10.1126/science.abb9811}
  {https://www.science.org/doi/pdf/10.1126/science.abb9811} \BibitemShut
  {NoStop}%
\bibitem [{\citenamefont {Mi}\ \emph {et~al.}(2021)\citenamefont {Mi},
  \citenamefont {Ippoliti}, \citenamefont {Quintana}, \citenamefont {Greene},
  \citenamefont {Chen}, \citenamefont {Gross}, \citenamefont {Arute},
  \citenamefont {Arya}, \citenamefont {Atalaya}, \citenamefont {Babbush},
  \citenamefont {Bardin}, \citenamefont {Basso}, \citenamefont {Bengtsson},
  \citenamefont {Bilmes}, \citenamefont {Bourassa}, \citenamefont {Brill},
  \citenamefont {Broughton}, \citenamefont {Buckley}, \citenamefont {Buell},
  \citenamefont {Burkett}, \citenamefont {Bushnell}, \citenamefont {Chiaro},
  \citenamefont {Collins}, \citenamefont {Courtney}, \citenamefont {Debroy},
  \citenamefont {Demura}, \citenamefont {Derk}, \citenamefont {Dunsworth},
  \citenamefont {Eppens}, \citenamefont {Erickson}, \citenamefont {Farhi},
  \citenamefont {Fowler}, \citenamefont {Foxen}, \citenamefont {Gidney},
  \citenamefont {Giustina}, \citenamefont {Harrigan}, \citenamefont
  {Harrington}, \citenamefont {Hilton}, \citenamefont {Ho}, \citenamefont
  {Hong}, \citenamefont {Huang}, \citenamefont {Huff}, \citenamefont {Huggins},
  \citenamefont {Ioffe}, \citenamefont {Isakov}, \citenamefont {Iveland},
  \citenamefont {Jeffrey}, \citenamefont {Jiang}, \citenamefont {Jones},
  \citenamefont {Kafri}, \citenamefont {Khattar}, \citenamefont {Kim},
  \citenamefont {Kitaev}, \citenamefont {Klimov}, \citenamefont {Korotkov},
  \citenamefont {Kostritsa}, \citenamefont {Landhuis}, \citenamefont {Laptev},
  \citenamefont {Lee}, \citenamefont {Lee}, \citenamefont {Locharla},
  \citenamefont {Lucero}, \citenamefont {Martin}, \citenamefont {McClean},
  \citenamefont {McCourt}, \citenamefont {McEwen}, \citenamefont {Miao},
  \citenamefont {Mohseni}, \citenamefont {Montazeri}, \citenamefont
  {Mruczkiewicz}, \citenamefont {Naaman}, \citenamefont {Neeley}, \citenamefont
  {Neill}, \citenamefont {Newman}, \citenamefont {Niu}, \citenamefont
  {O'Brien}, \citenamefont {Opremcak}, \citenamefont {Ostby}, \citenamefont
  {Pato}, \citenamefont {Petukhov}, \citenamefont {Rubin}, \citenamefont
  {Sank}, \citenamefont {Satzinger}, \citenamefont {Shvarts}, \citenamefont
  {Su}, \citenamefont {Strain}, \citenamefont {Szalay}, \citenamefont
  {Trevithick}, \citenamefont {Villalonga}, \citenamefont {White},
  \citenamefont {Yao}, \citenamefont {Yeh}, \citenamefont {Yoo}, \citenamefont
  {Zalcman}, \citenamefont {Neven}, \citenamefont {Boixo}, \citenamefont
  {Smelyanskiy}, \citenamefont {Megrant}, \citenamefont {Kelly}, \citenamefont
  {Chen}, \citenamefont {Sondhi}, \citenamefont {Moessner}, \citenamefont
  {Kechedzhi}, \citenamefont {Khemani},\ and\ \citenamefont
  {Roushan}}]{Time_Crystal}%
  \BibitemOpen
  \bibfield  {author} {\bibinfo {author} {\bibfnamefont {X.}~\bibnamefont
  {Mi}}, \bibinfo {author} {\bibfnamefont {M.}~\bibnamefont {Ippoliti}},
  \bibinfo {author} {\bibfnamefont {C.}~\bibnamefont {Quintana}}, \bibinfo
  {author} {\bibfnamefont {A.}~\bibnamefont {Greene}}, \bibinfo {author}
  {\bibfnamefont {Z.}~\bibnamefont {Chen}}, \bibinfo {author} {\bibfnamefont
  {J.}~\bibnamefont {Gross}}, \bibinfo {author} {\bibfnamefont
  {F.}~\bibnamefont {Arute}}, \bibinfo {author} {\bibfnamefont
  {K.}~\bibnamefont {Arya}}, \bibinfo {author} {\bibfnamefont {J.}~\bibnamefont
  {Atalaya}}, \bibinfo {author} {\bibfnamefont {R.}~\bibnamefont {Babbush}},
  \bibinfo {author} {\bibfnamefont {J.~C.}\ \bibnamefont {Bardin}}, \bibinfo
  {author} {\bibfnamefont {J.}~\bibnamefont {Basso}}, \bibinfo {author}
  {\bibfnamefont {A.}~\bibnamefont {Bengtsson}}, \bibinfo {author}
  {\bibfnamefont {A.}~\bibnamefont {Bilmes}}, \bibinfo {author} {\bibfnamefont
  {A.}~\bibnamefont {Bourassa}}, \bibinfo {author} {\bibfnamefont
  {L.}~\bibnamefont {Brill}}, \bibinfo {author} {\bibfnamefont
  {M.}~\bibnamefont {Broughton}}, \bibinfo {author} {\bibfnamefont {B.~B.}\
  \bibnamefont {Buckley}}, \bibinfo {author} {\bibfnamefont {D.~A.}\
  \bibnamefont {Buell}}, \bibinfo {author} {\bibfnamefont {B.}~\bibnamefont
  {Burkett}}, \bibinfo {author} {\bibfnamefont {N.}~\bibnamefont {Bushnell}},
  \bibinfo {author} {\bibfnamefont {B.}~\bibnamefont {Chiaro}}, \bibinfo
  {author} {\bibfnamefont {R.}~\bibnamefont {Collins}}, \bibinfo {author}
  {\bibfnamefont {W.}~\bibnamefont {Courtney}}, \bibinfo {author}
  {\bibfnamefont {D.}~\bibnamefont {Debroy}}, \bibinfo {author} {\bibfnamefont
  {S.}~\bibnamefont {Demura}}, \bibinfo {author} {\bibfnamefont {A.~R.}\
  \bibnamefont {Derk}}, \bibinfo {author} {\bibfnamefont {A.}~\bibnamefont
  {Dunsworth}}, \bibinfo {author} {\bibfnamefont {D.}~\bibnamefont {Eppens}},
  \bibinfo {author} {\bibfnamefont {C.}~\bibnamefont {Erickson}}, \bibinfo
  {author} {\bibfnamefont {E.}~\bibnamefont {Farhi}}, \bibinfo {author}
  {\bibfnamefont {A.~G.}\ \bibnamefont {Fowler}}, \bibinfo {author}
  {\bibfnamefont {B.}~\bibnamefont {Foxen}}, \bibinfo {author} {\bibfnamefont
  {C.}~\bibnamefont {Gidney}}, \bibinfo {author} {\bibfnamefont
  {M.}~\bibnamefont {Giustina}}, \bibinfo {author} {\bibfnamefont {M.~P.}\
  \bibnamefont {Harrigan}}, \bibinfo {author} {\bibfnamefont {S.~D.}\
  \bibnamefont {Harrington}}, \bibinfo {author} {\bibfnamefont
  {J.}~\bibnamefont {Hilton}}, \bibinfo {author} {\bibfnamefont
  {A.}~\bibnamefont {Ho}}, \bibinfo {author} {\bibfnamefont {S.}~\bibnamefont
  {Hong}}, \bibinfo {author} {\bibfnamefont {T.}~\bibnamefont {Huang}},
  \bibinfo {author} {\bibfnamefont {A.}~\bibnamefont {Huff}}, \bibinfo {author}
  {\bibfnamefont {W.~J.}\ \bibnamefont {Huggins}}, \bibinfo {author}
  {\bibfnamefont {L.~B.}\ \bibnamefont {Ioffe}}, \bibinfo {author}
  {\bibfnamefont {S.~V.}\ \bibnamefont {Isakov}}, \bibinfo {author}
  {\bibfnamefont {J.}~\bibnamefont {Iveland}}, \bibinfo {author} {\bibfnamefont
  {E.}~\bibnamefont {Jeffrey}}, \bibinfo {author} {\bibfnamefont
  {Z.}~\bibnamefont {Jiang}}, \bibinfo {author} {\bibfnamefont
  {C.}~\bibnamefont {Jones}}, \bibinfo {author} {\bibfnamefont
  {D.}~\bibnamefont {Kafri}}, \bibinfo {author} {\bibfnamefont
  {T.}~\bibnamefont {Khattar}}, \bibinfo {author} {\bibfnamefont
  {S.}~\bibnamefont {Kim}}, \bibinfo {author} {\bibfnamefont {A.}~\bibnamefont
  {Kitaev}}, \bibinfo {author} {\bibfnamefont {P.~V.}\ \bibnamefont {Klimov}},
  \bibinfo {author} {\bibfnamefont {A.~N.}\ \bibnamefont {Korotkov}}, \bibinfo
  {author} {\bibfnamefont {F.}~\bibnamefont {Kostritsa}}, \bibinfo {author}
  {\bibfnamefont {D.}~\bibnamefont {Landhuis}}, \bibinfo {author}
  {\bibfnamefont {P.}~\bibnamefont {Laptev}}, \bibinfo {author} {\bibfnamefont
  {J.}~\bibnamefont {Lee}}, \bibinfo {author} {\bibfnamefont {K.}~\bibnamefont
  {Lee}}, \bibinfo {author} {\bibfnamefont {A.}~\bibnamefont {Locharla}},
  \bibinfo {author} {\bibfnamefont {E.}~\bibnamefont {Lucero}}, \bibinfo
  {author} {\bibfnamefont {O.}~\bibnamefont {Martin}}, \bibinfo {author}
  {\bibfnamefont {J.~R.}\ \bibnamefont {McClean}}, \bibinfo {author}
  {\bibfnamefont {T.}~\bibnamefont {McCourt}}, \bibinfo {author} {\bibfnamefont
  {M.}~\bibnamefont {McEwen}}, \bibinfo {author} {\bibfnamefont {K.~C.}\
  \bibnamefont {Miao}}, \bibinfo {author} {\bibfnamefont {M.}~\bibnamefont
  {Mohseni}}, \bibinfo {author} {\bibfnamefont {S.}~\bibnamefont {Montazeri}},
  \bibinfo {author} {\bibfnamefont {W.}~\bibnamefont {Mruczkiewicz}}, \bibinfo
  {author} {\bibfnamefont {O.}~\bibnamefont {Naaman}}, \bibinfo {author}
  {\bibfnamefont {M.}~\bibnamefont {Neeley}}, \bibinfo {author} {\bibfnamefont
  {C.}~\bibnamefont {Neill}}, \bibinfo {author} {\bibfnamefont
  {M.}~\bibnamefont {Newman}}, \bibinfo {author} {\bibfnamefont {M.~Y.}\
  \bibnamefont {Niu}}, \bibinfo {author} {\bibfnamefont {T.~E.}\ \bibnamefont
  {O'Brien}}, \bibinfo {author} {\bibfnamefont {A.}~\bibnamefont {Opremcak}},
  \bibinfo {author} {\bibfnamefont {E.}~\bibnamefont {Ostby}}, \bibinfo
  {author} {\bibfnamefont {B.}~\bibnamefont {Pato}}, \bibinfo {author}
  {\bibfnamefont {A.}~\bibnamefont {Petukhov}}, \bibinfo {author}
  {\bibfnamefont {N.~C.}\ \bibnamefont {Rubin}}, \bibinfo {author}
  {\bibfnamefont {D.}~\bibnamefont {Sank}}, \bibinfo {author} {\bibfnamefont
  {K.~J.}\ \bibnamefont {Satzinger}}, \bibinfo {author} {\bibfnamefont
  {V.}~\bibnamefont {Shvarts}}, \bibinfo {author} {\bibfnamefont
  {Y.}~\bibnamefont {Su}}, \bibinfo {author} {\bibfnamefont {D.}~\bibnamefont
  {Strain}}, \bibinfo {author} {\bibfnamefont {M.}~\bibnamefont {Szalay}},
  \bibinfo {author} {\bibfnamefont {M.~D.}\ \bibnamefont {Trevithick}},
  \bibinfo {author} {\bibfnamefont {B.}~\bibnamefont {Villalonga}}, \bibinfo
  {author} {\bibfnamefont {T.}~\bibnamefont {White}}, \bibinfo {author}
  {\bibfnamefont {Z.~J.}\ \bibnamefont {Yao}}, \bibinfo {author} {\bibfnamefont
  {P.}~\bibnamefont {Yeh}}, \bibinfo {author} {\bibfnamefont {J.}~\bibnamefont
  {Yoo}}, \bibinfo {author} {\bibfnamefont {A.}~\bibnamefont {Zalcman}},
  \bibinfo {author} {\bibfnamefont {H.}~\bibnamefont {Neven}}, \bibinfo
  {author} {\bibfnamefont {S.}~\bibnamefont {Boixo}}, \bibinfo {author}
  {\bibfnamefont {V.}~\bibnamefont {Smelyanskiy}}, \bibinfo {author}
  {\bibfnamefont {A.}~\bibnamefont {Megrant}}, \bibinfo {author} {\bibfnamefont
  {J.}~\bibnamefont {Kelly}}, \bibinfo {author} {\bibfnamefont
  {Y.}~\bibnamefont {Chen}}, \bibinfo {author} {\bibfnamefont {S.~L.}\
  \bibnamefont {Sondhi}}, \bibinfo {author} {\bibfnamefont {R.}~\bibnamefont
  {Moessner}}, \bibinfo {author} {\bibfnamefont {K.}~\bibnamefont {Kechedzhi}},
  \bibinfo {author} {\bibfnamefont {V.}~\bibnamefont {Khemani}},\ and\ \bibinfo
  {author} {\bibfnamefont {P.}~\bibnamefont {Roushan}},\ }\bibfield  {title}
  {\bibinfo {title} {Time-crystalline eigenstate order on a quantum
  processor},\ }\href {https://doi.org/10.1038/s41586-021-04257-w} {\bibfield
  {journal} {\bibinfo  {journal} {Nature}\ }\textbf {\bibinfo {volume} {601}},\
  \bibinfo {pages} {531} (\bibinfo {year} {2021})}\BibitemShut {NoStop}%
\bibitem [{\citenamefont {Bauer}\ \emph {et~al.}(2021)\citenamefont {Bauer},
  \citenamefont {Nachman},\ and\ \citenamefont {Freytsis}}]{high_energy}%
  \BibitemOpen
  \bibfield  {author} {\bibinfo {author} {\bibfnamefont {C.~W.}\ \bibnamefont
  {Bauer}}, \bibinfo {author} {\bibfnamefont {B.}~\bibnamefont {Nachman}},\
  and\ \bibinfo {author} {\bibfnamefont {M.}~\bibnamefont {Freytsis}},\
  }\bibfield  {title} {\bibinfo {title} {Simulating collider physics on quantum
  computers using effective field theories},\ }\bibfield  {journal} {\bibinfo
  {journal} {Physical Review Letters}\ }\textbf {\bibinfo {volume} {127}},\
  \href {https://doi.org/10.1103/physrevlett.127.212001}
  {10.1103/physrevlett.127.212001} (\bibinfo {year} {2021})\BibitemShut
  {NoStop}%
\bibitem [{\citenamefont {Fedorov}\ \emph {et~al.}(2022)\citenamefont
  {Fedorov}, \citenamefont {Peng},\ and\ \citenamefont {Govind}}]{Scalability}%
  \BibitemOpen
  \bibfield  {author} {\bibinfo {author} {\bibfnamefont {D.}~\bibnamefont
  {Fedorov}}, \bibinfo {author} {\bibfnamefont {B.}~\bibnamefont {Peng}},\ and\
  \bibinfo {author} {\bibfnamefont {N.~e.~a.}\ \bibnamefont {Govind}},\
  }\bibfield  {title} {\bibinfo {title} {Vqe method: a short survey and recent
  developments},\ }\bibfield  {journal} {\bibinfo  {journal} {Materials
  Theory}\ }\textbf {\bibinfo {volume} {6}},\ \href
  {https://doi.org/10.1186/s41313-021-00032-6} {10.1186/s41313-021-00032-6}
  (\bibinfo {year} {2022})\BibitemShut {NoStop}%
\bibitem [{\citenamefont {Peruzzo}\ \emph {et~al.}(2018)\citenamefont
  {Peruzzo}, \citenamefont {McClean}, \citenamefont {Shadbolt}, \citenamefont
  {Yung}, \citenamefont {Zhou},\ and\ \citenamefont {Love}}]{barren_plateau}%
  \BibitemOpen
  \bibfield  {author} {\bibinfo {author} {\bibfnamefont {A.}~\bibnamefont
  {Peruzzo}}, \bibinfo {author} {\bibfnamefont {J.}~\bibnamefont {McClean}},
  \bibinfo {author} {\bibfnamefont {P.}~\bibnamefont {Shadbolt}}, \bibinfo
  {author} {\bibfnamefont {M.~H.}\ \bibnamefont {Yung}}, \bibinfo {author}
  {\bibfnamefont {X.~Q.}\ \bibnamefont {Zhou}},\ and\ \bibinfo {author}
  {\bibfnamefont {P.~J.}\ \bibnamefont {Love}},\ }\bibfield  {title} {\bibinfo
  {title} {Barren plateaus in quantum neural network training landscapes},\
  }\href {https://doi.org/10.1038/s41467-018-07090-4} {\bibfield  {journal}
  {\bibinfo  {journal} {Nature Communications}\ }\textbf {\bibinfo {volume}
  {5}},\ \bibinfo {pages} {4812} (\bibinfo {year} {2018})}\BibitemShut
  {NoStop}%
\bibitem [{\citenamefont {Kim}\ and\ \citenamefont
  {Oz}(2021)}]{entanglement_barren}%
  \BibitemOpen
  \bibfield  {author} {\bibinfo {author} {\bibfnamefont {J.}~\bibnamefont
  {Kim}}\ and\ \bibinfo {author} {\bibfnamefont {Y.}~\bibnamefont {Oz}},\
  }\href {https://doi.org/10.48550/ARXIV.2102.12534} {\bibinfo {title}
  {Entanglement diagnostics for efficient quantum computation}} (\bibinfo
  {year} {2021})\BibitemShut {NoStop}%
\bibitem [{\citenamefont {Holmes}\ \emph {et~al.}(2022)\citenamefont {Holmes},
  \citenamefont {Sharma}, \citenamefont {Cerezo},\ and\ \citenamefont
  {Coles}}]{Holmes_2022}%
  \BibitemOpen
  \bibfield  {author} {\bibinfo {author} {\bibfnamefont {Z.}~\bibnamefont
  {Holmes}}, \bibinfo {author} {\bibfnamefont {K.}~\bibnamefont {Sharma}},
  \bibinfo {author} {\bibfnamefont {M.}~\bibnamefont {Cerezo}},\ and\ \bibinfo
  {author} {\bibfnamefont {P.~J.}\ \bibnamefont {Coles}},\ }\bibfield  {title}
  {\bibinfo {title} {Connecting ansatz expressibility to gradient magnitudes
  and barren plateaus},\ }\bibfield  {journal} {\bibinfo  {journal} {{PRX}
  Quantum}\ }\textbf {\bibinfo {volume} {3}},\ \href
  {https://doi.org/10.1103/prxquantum.3.010313} {10.1103/prxquantum.3.010313}
  (\bibinfo {year} {2022})\BibitemShut {NoStop}%
\bibitem [{\citenamefont {Aaronson}\ and\ \citenamefont
  {Gottesman}(2004)}]{Aaronson_2004}%
  \BibitemOpen
  \bibfield  {author} {\bibinfo {author} {\bibfnamefont {S.}~\bibnamefont
  {Aaronson}}\ and\ \bibinfo {author} {\bibfnamefont {D.}~\bibnamefont
  {Gottesman}},\ }\bibfield  {title} {\bibinfo {title} {Improved simulation of
  stabilizer circuits},\ }\bibfield  {journal} {\bibinfo  {journal} {Physical
  Review A}\ }\textbf {\bibinfo {volume} {70}},\ \href
  {https://doi.org/10.1103/physreva.70.052328} {10.1103/physreva.70.052328}
  (\bibinfo {year} {2004})\BibitemShut {NoStop}%
\bibitem [{\citenamefont {Ravi}\ \emph {et~al.}(2022)\citenamefont {Ravi},
  \citenamefont {Gokhale}, \citenamefont {Ding}, \citenamefont {Kirby},
  \citenamefont {Smith}, \citenamefont {Baker}, \citenamefont {Love},
  \citenamefont {Hoffmann}, \citenamefont {Brown},\ and\ \citenamefont
  {Chong}}]{CAFQA}%
  \BibitemOpen
  \bibfield  {author} {\bibinfo {author} {\bibfnamefont {G.~S.}\ \bibnamefont
  {Ravi}}, \bibinfo {author} {\bibfnamefont {P.}~\bibnamefont {Gokhale}},
  \bibinfo {author} {\bibfnamefont {Y.}~\bibnamefont {Ding}}, \bibinfo {author}
  {\bibfnamefont {W.~M.}\ \bibnamefont {Kirby}}, \bibinfo {author}
  {\bibfnamefont {K.~N.}\ \bibnamefont {Smith}}, \bibinfo {author}
  {\bibfnamefont {J.~M.}\ \bibnamefont {Baker}}, \bibinfo {author}
  {\bibfnamefont {P.~J.}\ \bibnamefont {Love}}, \bibinfo {author}
  {\bibfnamefont {H.}~\bibnamefont {Hoffmann}}, \bibinfo {author}
  {\bibfnamefont {K.~R.}\ \bibnamefont {Brown}},\ and\ \bibinfo {author}
  {\bibfnamefont {F.~T.}\ \bibnamefont {Chong}},\ }\href
  {https://doi.org/10.48550/ARXIV.2202.12924} {\bibinfo {title} {Cafqa:
  Clifford ansatz for quantum accuracy}} (\bibinfo {year} {2022})\BibitemShut
  {NoStop}%
\bibitem [{\citenamefont {Mitarai}\ \emph {et~al.}(2018)\citenamefont
  {Mitarai}, \citenamefont {Negoro}, \citenamefont {Kitagawa},\ and\
  \citenamefont {Fujii}}]{parametershift}%
  \BibitemOpen
  \bibfield  {author} {\bibinfo {author} {\bibfnamefont {K.}~\bibnamefont
  {Mitarai}}, \bibinfo {author} {\bibfnamefont {M.}~\bibnamefont {Negoro}},
  \bibinfo {author} {\bibfnamefont {M.}~\bibnamefont {Kitagawa}},\ and\
  \bibinfo {author} {\bibfnamefont {K.}~\bibnamefont {Fujii}},\ }\bibfield
  {title} {\bibinfo {title} {Quantum circuit learning},\ }\bibfield  {journal}
  {\bibinfo  {journal} {Physical Review A}\ }\textbf {\bibinfo {volume} {98}},\
  \href {https://doi.org/10.1103/physreva.98.032309}
  {10.1103/physreva.98.032309} (\bibinfo {year} {2018})\BibitemShut {NoStop}%
\bibitem [{\citenamefont {Grant}\ \emph {et~al.}(2019)\citenamefont {Grant},
  \citenamefont {Wossnig}, \citenamefont {Ostaszewski},\ and\ \citenamefont
  {Benedetti}}]{Initialisation_1}%
  \BibitemOpen
  \bibfield  {author} {\bibinfo {author} {\bibfnamefont {E.}~\bibnamefont
  {Grant}}, \bibinfo {author} {\bibfnamefont {L.}~\bibnamefont {Wossnig}},
  \bibinfo {author} {\bibfnamefont {M.}~\bibnamefont {Ostaszewski}},\ and\
  \bibinfo {author} {\bibfnamefont {M.}~\bibnamefont {Benedetti}},\ }\bibfield
  {title} {\bibinfo {title} {An initialization strategy for addressing barren
  plateaus in parametrized quantum circuits},\ }\href
  {https://doi.org/10.22331/q-2019-12-09-214} {\bibfield  {journal} {\bibinfo
  {journal} {{Quantum}}\ }\textbf {\bibinfo {volume} {3}},\ \bibinfo {pages}
  {214} (\bibinfo {year} {2019})}\BibitemShut {NoStop}%
\bibitem [{\citenamefont {Dborin}\ \emph {et~al.}(2021)\citenamefont {Dborin},
  \citenamefont {Barratt}, \citenamefont {Wimalaweera}, \citenamefont
  {Wright},\ and\ \citenamefont {Green}}]{Initialisation_2}%
  \BibitemOpen
  \bibfield  {author} {\bibinfo {author} {\bibfnamefont {J.}~\bibnamefont
  {Dborin}}, \bibinfo {author} {\bibfnamefont {F.}~\bibnamefont {Barratt}},
  \bibinfo {author} {\bibfnamefont {V.}~\bibnamefont {Wimalaweera}}, \bibinfo
  {author} {\bibfnamefont {L.}~\bibnamefont {Wright}},\ and\ \bibinfo {author}
  {\bibfnamefont {A.~G.}\ \bibnamefont {Green}},\ }\href
  {https://doi.org/10.48550/ARXIV.2106.05742} {\bibinfo {title} {Matrix product
  state pre-training for quantum machine learning}} (\bibinfo {year}
  {2021})\BibitemShut {NoStop}%
\bibitem [{\citenamefont {Skolik}\ \emph {et~al.}(2021)\citenamefont {Skolik},
  \citenamefont {McClean}, \citenamefont {Mohseni}, \citenamefont {van~der
  Smagt},\ and\ \citenamefont {Leib}}]{Initialisation_3}%
  \BibitemOpen
  \bibfield  {author} {\bibinfo {author} {\bibfnamefont {A.}~\bibnamefont
  {Skolik}}, \bibinfo {author} {\bibfnamefont {J.~R.}\ \bibnamefont {McClean}},
  \bibinfo {author} {\bibfnamefont {M.}~\bibnamefont {Mohseni}}, \bibinfo
  {author} {\bibfnamefont {P.}~\bibnamefont {van~der Smagt}},\ and\ \bibinfo
  {author} {\bibfnamefont {M.}~\bibnamefont {Leib}},\ }\bibfield  {title}
  {\bibinfo {title} {Layerwise learning for quantum neural networks},\
  }\bibfield  {journal} {\bibinfo  {journal} {Quantum Machine Intelligence}\
  }\textbf {\bibinfo {volume} {3}},\ \href
  {https://doi.org/10.1007/s42484-020-00036-4} {10.1007/s42484-020-00036-4}
  (\bibinfo {year} {2021})\BibitemShut {NoStop}%
\bibitem [{\citenamefont {Gottesman}(1998)}]{Gottesman}%
  \BibitemOpen
  \bibfield  {author} {\bibinfo {author} {\bibfnamefont {D.}~\bibnamefont
  {Gottesman}},\ }\bibfield  {title} {\bibinfo {title} {The heisenberg
  representation of quantum computers}\ }\href
  {https://doi.org/10.48550/ARXIV.QUANT-PH/9807006}
  {10.48550/ARXIV.QUANT-PH/9807006} (\bibinfo {year} {1998})\BibitemShut
  {NoStop}%
\bibitem [{\citenamefont {Anand}\ \emph {et~al.}(2022)\citenamefont {Anand},
  \citenamefont {Schleich}, \citenamefont {Alperin-Lea}, \citenamefont
  {Jensen}, \citenamefont {Sim}, \citenamefont {D{\'{\i} }az-Tinoco},
  \citenamefont {Kottmann}, \citenamefont {Degroote}, \citenamefont
  {Izmaylov},\ and\ \citenamefont {Aspuru-Guzik}}]{UCCSD}%
  \BibitemOpen
  \bibfield  {author} {\bibinfo {author} {\bibfnamefont {A.}~\bibnamefont
  {Anand}}, \bibinfo {author} {\bibfnamefont {P.}~\bibnamefont {Schleich}},
  \bibinfo {author} {\bibfnamefont {S.}~\bibnamefont {Alperin-Lea}}, \bibinfo
  {author} {\bibfnamefont {P.~W.~K.}\ \bibnamefont {Jensen}}, \bibinfo {author}
  {\bibfnamefont {S.}~\bibnamefont {Sim}}, \bibinfo {author} {\bibfnamefont
  {M.}~\bibnamefont {D{\'{\i} }az-Tinoco}}, \bibinfo {author} {\bibfnamefont
  {J.~S.}\ \bibnamefont {Kottmann}}, \bibinfo {author} {\bibfnamefont
  {M.}~\bibnamefont {Degroote}}, \bibinfo {author} {\bibfnamefont {A.~F.}\
  \bibnamefont {Izmaylov}},\ and\ \bibinfo {author} {\bibfnamefont
  {A.}~\bibnamefont {Aspuru-Guzik}},\ }\bibfield  {title} {\bibinfo {title} {A
  quantum computing view on unitary coupled cluster theory},\ }\href
  {https://doi.org/10.1039/d1cs00932j} {\bibfield  {journal} {\bibinfo
  {journal} {Chemical Society Reviews}\ }\textbf {\bibinfo {volume} {51}},\
  \bibinfo {pages} {1659} (\bibinfo {year} {2022})}\BibitemShut {NoStop}%
\bibitem [{\citenamefont {Harrow}\ and\ \citenamefont {Low}(2009)}]{2-design}%
  \BibitemOpen
  \bibfield  {author} {\bibinfo {author} {\bibfnamefont {A.~W.}\ \bibnamefont
  {Harrow}}\ and\ \bibinfo {author} {\bibfnamefont {R.~A.}\ \bibnamefont
  {Low}},\ }\bibfield  {title} {\bibinfo {title} {Random quantum circuits are
  approximate 2-designs},\ }\href {https://doi.org/10.1007/s00220-009-0873-6}
  {\bibfield  {journal} {\bibinfo  {journal} {Communications in Mathematical
  Physics}\ }\textbf {\bibinfo {volume} {291}},\ \bibinfo {pages} {257}
  (\bibinfo {year} {2009})}\BibitemShut {NoStop}%
\bibitem [{\citenamefont {Fradkin}(1989)}]{Jordan-Wigner}%
  \BibitemOpen
  \bibfield  {author} {\bibinfo {author} {\bibfnamefont {E.}~\bibnamefont
  {Fradkin}},\ }\bibfield  {title} {\bibinfo {title} {Jordan-wigner
  transformation for quantum-spin systems in two dimensions and fractional
  statistics},\ }\href {https://doi.org/10.1103/PhysRevLett.63.322} {\bibfield
  {journal} {\bibinfo  {journal} {Phys. Rev. Lett.}\ }\textbf {\bibinfo
  {volume} {63}},\ \bibinfo {pages} {322} (\bibinfo {year} {1989})}\BibitemShut
  {NoStop}%
\bibitem [{\citenamefont {Seeley}\ \emph {et~al.}(2012)\citenamefont {Seeley},
  \citenamefont {Richard},\ and\ \citenamefont {Love}}]{Parity}%
  \BibitemOpen
  \bibfield  {author} {\bibinfo {author} {\bibfnamefont {J.~T.}\ \bibnamefont
  {Seeley}}, \bibinfo {author} {\bibfnamefont {M.~J.}\ \bibnamefont
  {Richard}},\ and\ \bibinfo {author} {\bibfnamefont {P.~J.}\ \bibnamefont
  {Love}},\ }\bibfield  {title} {\bibinfo {title} {The bravyi-kitaev
  transformation for quantum computation of electronic structure},\ }\href
  {https://doi.org/10.1063/1.4768229} {\bibfield  {journal} {\bibinfo
  {journal} {The Journal of Chemical Physics}\ }\textbf {\bibinfo {volume}
  {137}},\ \bibinfo {pages} {224109} (\bibinfo {year} {2012})}\BibitemShut
  {NoStop}%
\bibitem [{\citenamefont {Bravyi}\ and\ \citenamefont {Kitaev}(2002)}]{Bravyi}%
  \BibitemOpen
  \bibfield  {author} {\bibinfo {author} {\bibfnamefont {S.~B.}\ \bibnamefont
  {Bravyi}}\ and\ \bibinfo {author} {\bibfnamefont {A.~Y.}\ \bibnamefont
  {Kitaev}},\ }\bibfield  {title} {\bibinfo {title} {Fermionic quantum
  computation},\ }\href {https://doi.org/10.1006/aphy.2002.6254} {\bibfield
  {journal} {\bibinfo  {journal} {Annals of Physics}\ }\textbf {\bibinfo
  {volume} {298}},\ \bibinfo {pages} {210} (\bibinfo {year}
  {2002})}\BibitemShut {NoStop}%
\bibitem [{\citenamefont {Kempe}\ \emph {et~al.}(2004)\citenamefont {Kempe},
  \citenamefont {Kitaev},\ and\ \citenamefont {Regev}}]{k-local}%
  \BibitemOpen
  \bibfield  {author} {\bibinfo {author} {\bibfnamefont {J.}~\bibnamefont
  {Kempe}}, \bibinfo {author} {\bibfnamefont {A.}~\bibnamefont {Kitaev}},\ and\
  \bibinfo {author} {\bibfnamefont {O.}~\bibnamefont {Regev}},\ }\bibfield
  {title} {\bibinfo {title} {The complexity of the local hamiltonian problem}\
  }\href {https://doi.org/10.48550/ARXIV.QUANT-PH/0406180}
  {10.48550/ARXIV.QUANT-PH/0406180} (\bibinfo {year} {2004})\BibitemShut
  {NoStop}%
\bibitem [{\citenamefont {Guerreschi}(2021)}]{QUBO_Ising}%
  \BibitemOpen
  \bibfield  {author} {\bibinfo {author} {\bibfnamefont {G.~G.}\ \bibnamefont
  {Guerreschi}},\ }\href {https://doi.org/10.48550/ARXIV.2101.07813} {\bibinfo
  {title} {Solving quadratic unconstrained binary optimization with
  divide-and-conquer and quantum algorithms}} (\bibinfo {year}
  {2021})\BibitemShut {NoStop}%
\bibitem [{\citenamefont {Dutta}\ \emph {et~al.}(2015)\citenamefont {Dutta},
  \citenamefont {Aeppli}, \citenamefont {Chakrabarti}, \citenamefont
  {Divakaran}, \citenamefont {Rosenbaum},\ and\ \citenamefont
  {Sen}}]{Volume_law}%
  \BibitemOpen
  \bibfield  {author} {\bibinfo {author} {\bibfnamefont {A.}~\bibnamefont
  {Dutta}}, \bibinfo {author} {\bibfnamefont {G.}~\bibnamefont {Aeppli}},
  \bibinfo {author} {\bibfnamefont {B.~K.}\ \bibnamefont {Chakrabarti}},
  \bibinfo {author} {\bibfnamefont {U.}~\bibnamefont {Divakaran}}, \bibinfo
  {author} {\bibfnamefont {T.~F.}\ \bibnamefont {Rosenbaum}},\ and\ \bibinfo
  {author} {\bibfnamefont {D.}~\bibnamefont {Sen}},\ }\href@noop {} {\emph
  {\bibinfo {title} {Quantum phase transitions in transverse field spin models:
  from statistical physics to quantum information}}}\ (\bibinfo  {publisher}
  {Cambridge University Press},\ \bibinfo {year} {2015})\BibitemShut {NoStop}%
\bibitem [{Note1()}]{Note1}%
  \BibitemOpen
  \bibinfo {note} {A low energy $\protect \ensuremath {\left \langle Z_iZ_j
  \right \rangle }$ stabilizer state is either an odd parity logical state (in
  $i,j$) in which case $\protect \ensuremath {\left \langle Z_i \right \rangle
  } =\pm 1 = -\protect \ensuremath {\left \langle Z_j \right \rangle }$, or a
  Bell-like state (in $i,j$) in which case $\protect \ensuremath {\left \langle
  Z_i \right \rangle } = \protect \ensuremath {\left \langle Z_j \right \rangle
  } = 0$}\BibitemShut {NoStop}%
\bibitem [{Note2()}]{Note2}%
  \BibitemOpen
  \bibinfo {note} {Here local minima is referring to the $\theta $ vector
  locality, not Hilbert-space locality as change in a gate angle can have a
  large change of the parameterised state in the Hilbert space}\BibitemShut
  {NoStop}%
\end{thebibliography}%

	\newpage    
	\mbox{}
	\newpage
	\FloatBarrier
	\appendix
	\widetext
	\section{Clifford circuit annealing} 
	\label{section: numericals}
\begin{figure}
		\centering
		\includegraphics[width = 0.75\columnwidth]{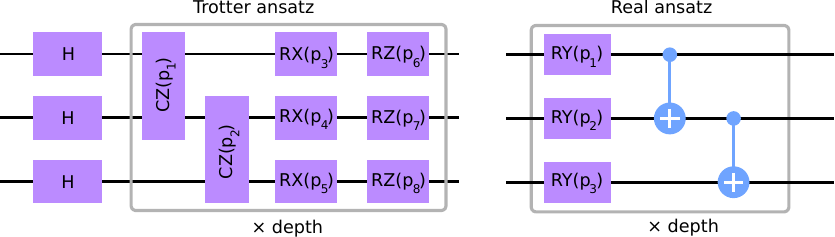}
		\caption{Ansatz circuits used for TFIM and quantum chemistry problems. Trotterised inspired ansatz (with independent parameters for each control-Z and single qubit rotations in each layer)  allows complex valued wave functions. The real-valued ansatz has only Y-rotations and control not gates and ensures the final wave function is real valued.}
		\label{fig:AnsatzDesign}
	\end{figure}

	Here we provide a detailed description of the Clifford pre-optimisation algorithm. The ansatz circuits we consider are (i) Complex valued ansatz that has the same structure as a Trotterized dynamics of an Ising model, and (ii) Real valued ansatz consisting of $CX$ and parameterised $R_y$ only, see Fig.~\ref{fig:AnsatzDesign}. Note that for a given depth $D$ and qubit count $n$ the Trotter ansatz as $D(3n-1)$ parameters while the Real ansatz has $Dn$. Each parameter is an $R_x,~R_z$ or $CZ$ (controlled Z) rotation. The algorithm considers the classically simulable sector of an ansatz quantum circuit, i.e. restricting parameterised rotations in the parameter vector $\theta = (\theta_1, \theta_2,... \theta_N)$ to each be multiples of $\pi/2$. 
	
	To minimize the state over this discrete set we consider classical annealing in the Clifford angles. This method randomly changes exactly two components (chosen heuristically) of the parameter vector $\theta$  --- keeping the changed parameters in $\{0, \pi/2, \pi, 3\pi/2\}$ --- to give a new $\theta'$. If the new cost $\Sij{\psi(\theta')}{H}{\psi(\theta')}$ is lower than the previous cost $\Sij{\psi(\theta)}{H}{\psi(\theta)}$ , the updated $\theta'$ vector is accepted with probability one, otherwise the new $\theta'$ is accepted with probability $\exp(-\beta \Delta E)$ with $\Delta E = \Sij{\psi(\theta')}{H}{\psi(\theta')} -\Sij{\psi(\theta)}{H}{\psi(\theta)}$. Here $\beta$ is the effective inverse temperature and is set to be on the order of the (estimated) characteristic energy gap between low energy states. If this gap is unknown then beta may be chosen to be arbitrarily large corresponding to zero temperature annealing. 
	
	To stop the annealing algorithm getting stuck in a local minima~\footnote{Here local minima is referring to the $\theta$ vector locality, not Hilbert-space locality as change in a gate angle can have a large change of the parameterised state in the Hilbert space} we randomly reset the full $\theta$ vector if the annealing hasn't found a better $\theta'$ point over the last $k$ iterations. For all plots we heuristically choose $k$ in the range of 400-1000. We note other annealing methods may be applicable, e.g. parallel tempering, or different (perhaps iteration-dependent) rules for selecting which individual $\theta_i$ parameters to update. Or other discrete optimisation algorithms, such as Bayesian optimisation~\cite{CAFQA}.

\begin{figure*}
		\centering
		\includegraphics[width =\columnwidth]{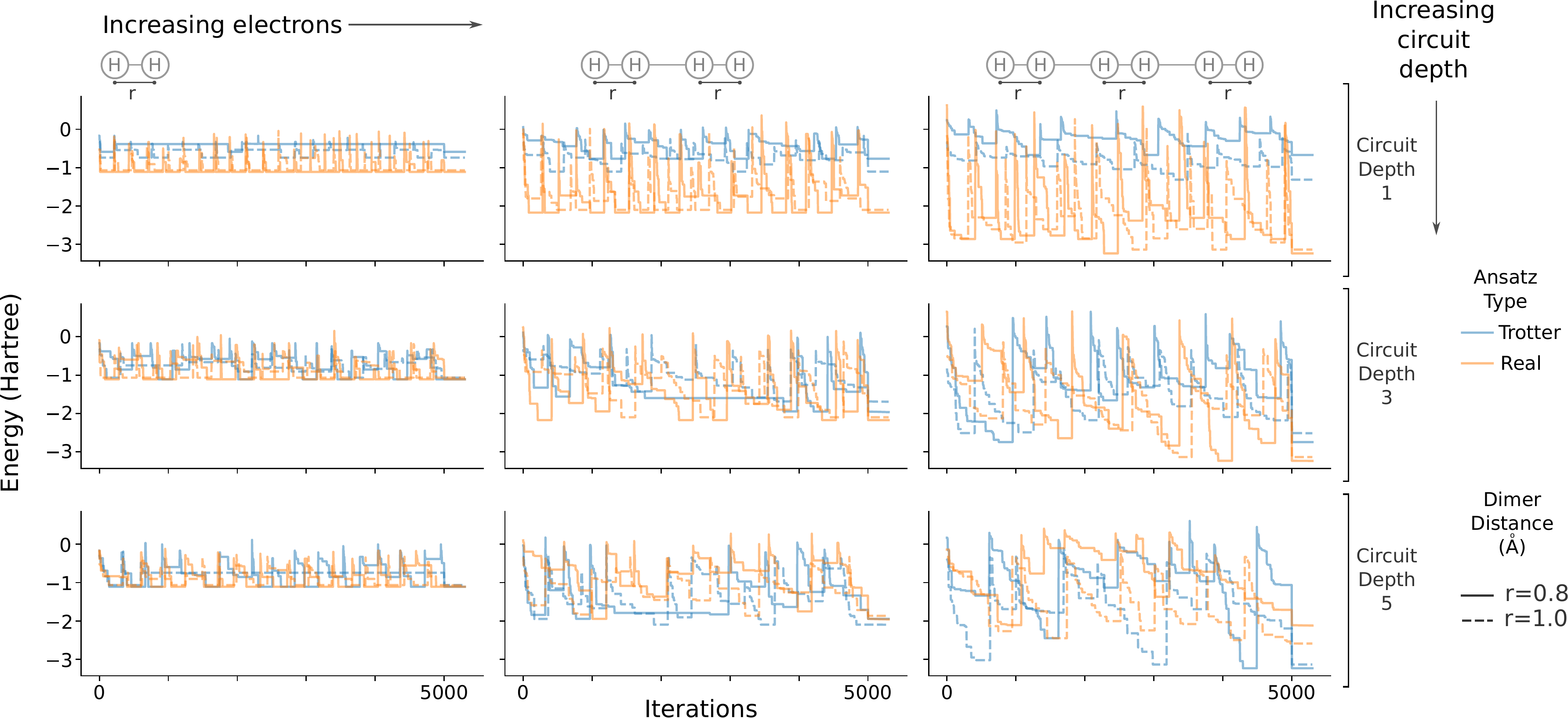}
		\caption{Plots shows ansatz circuit selection trajectories for three different quantum chemistry problems: (left) H$_2$ (middle) H$_4$ and (right) H$_6$, corresponding to four, six and eight qubits respectively (not including symmetry reduction). Rows correspond to different depth ansatz circuits, (top) depth one, (middle) depth three and (bottom) depth five, with the colors corresponding to the Trotter ansatz (blue) and real ansatz (orange). Line styles correspond to different dimer distances with a fixed inter dymer distance. Each curve is efficiently classically calculated and the set can be used to select an appropriate circuit for the on-device problem. Note the depth one real ansatz seems to find as good a solution as the depth five Trotter ansatz, but due to the few parameter, tends to find a good solution faster, and more consistently.}
		\label{fig:NH_timeseries}
	\end{figure*}

	The annealing trajectory is local in the parameter space (vector space of $\theta$), but can be non-local in terms of Hilbert space jumps of the resulting stabilizer state. There trajectory can therefore be dependent on the initial choice of the Clifford point. That can lead to initial condition-dependent local minima in the classical optimisation process. We note that our resetting after $k$ sequential non-improving evaluations helps prevent this issue. An example annealing iteration series in Fig. \ref{fig:NH_timeseries} shows the utility of threshold-dependent resetting in finding a good initial solution.

	\FloatBarrier
	\section{Clifford solution vs true ground states}
\begin{figure*}
		\centering
		\includegraphics[width =.8\columnwidth]{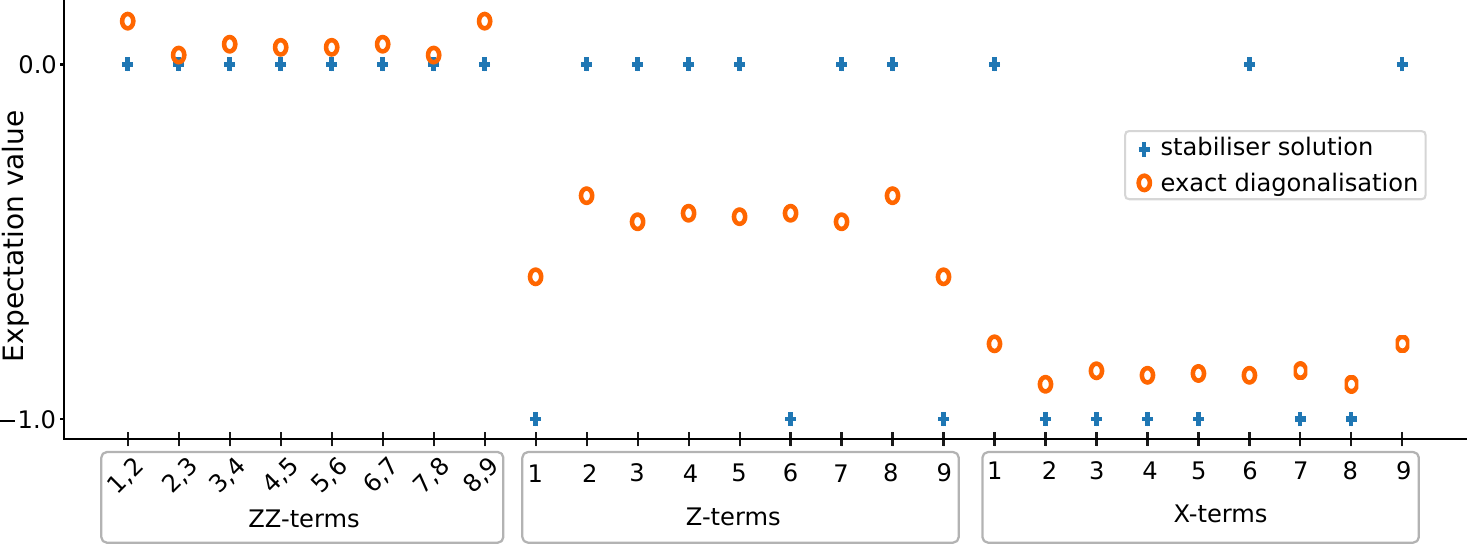}
		\caption{Expectation values of each individual Pauli string in a nine qubit TFIM Hamiltonian $H = \sum_i (Z_iZ_{i+1}) +2Z_i + 2X_i$, comparing the exact expectation values (orange circles) and  stabilizer solution (blue crosses). The stabilizer solution (constrained to be $\pm 1$ or zero) gets the general pattern of low/high correct, but with an emphasis on the $X_i$ stabilizer states as discussed in the main text.}
		\label{fig:H10_eig_1}
	\end{figure*}

	Other than observing the exact ground state energy, we can compare the stabilizer state and the ground state solution through looking at their respective contribution from individual Pauli strings in the Hamiltonian. The set of expectation values also tells us about the amount of magic in the exact ground state solution, since a state with high magic will have expectation values that deviate from $1, 0, -1$. If we observe a low magic ground state, it means that there is a stabilizer state that is close to the exact solution, which is good for the gradient descent. In the case of a high magic ground states (see Fig.~\ref{fig:H10_eig_1}), we can still find a stabilizer state with relatively low expectation value compared to random parameterisation. However, since the ground state solution is far from a  stabilzier state, the stabilizer state found is not guaranteed to be close to ground state, and still may be difficult to optimise.. 

\end{document}